\documentclass[conference]{IEEEtran}
\IEEEoverridecommandlockouts
\usepackage{xcolor}
\usepackage{textcomp}

\usepackage[absolute]{textpos}

\usepackage[most]{tcolorbox}
\usepackage{cite}
\usepackage{amsmath,amssymb,amsfonts}
\usepackage{algorithmic}
\usepackage{enumitem}
\usepackage{graphicx}
\usepackage{textcomp}
\usepackage{xcolor}
\usepackage{algorithm}
\usepackage{algorithmic}
\usepackage{comment}
\usepackage{listings}
\usepackage{xcolor}
\usepackage{listings}
\usepackage{color}
\usepackage{verbatim}
\usepackage{hyperref}
\usepackage{comment}
\usepackage{multirow}
\usepackage{tabularx}
\usepackage{booktabs}
\usepackage{colortbl}
\usepackage{multirow}
\usepackage{tabularx}
\usepackage{soul}
\usepackage{mathtools}
\usepackage{array}
\newcolumntype{P}[1]{>{\centering\hspace{0pt}\arraybackslash}p{#1}}
\newcolumntype{M}[1]{>{\centering\hspace{0pt}\arraybackslash}m{#1}}
\newcolumntype{C}[1]{>{\centering\arraybackslash}p{#1}}
\usepackage{pifont}

\definecolor{dkgreen}{rgb}{0,0.6,0}
\definecolor{gray}{rgb}{0.5,0.5,0.5}
\definecolor{mauve}{rgb}{0.58,0,0.82}

\lstdefinestyle{mystyle}{
    backgroundcolor=\color{white},   
    commentstyle=\color{codegreen},
    keywordstyle=\color{blue},
    numberstyle=\tiny\color{black},
    stringstyle=\color{purple},
    basicstyle=\ttfamily\scriptsize,
    breakatwhitespace=false,         
    breaklines=false,                 
    captionpos=b,                    
    keepspaces=false,                 
    numbers=left, 
    numbersep=3pt,                  
    showspaces=false,                
    showstringspaces=false,
    showtabs=false,                  
    tabsize=2,
    moredelim=**[is][\color{red}]{&}{&},
    moredelim=**[is][\color{myblue}]{/}{/}
}
\usepackage{xcolor}
\lstdefinestyle{mystyle2}{
    backgroundcolor=\color{blue!10!white},   
    commentstyle=\color{codegreen},
    keywordstyle=\color{blue},
    numberstyle=\tiny\color{black},
    stringstyle=\color{purple},
    basicstyle=\ttfamily\scriptsize,
    breakatwhitespace=false,         
    breaklines=false,                 
    captionpos=b,                    
    keepspaces=false,                 
    numbers=left, 
    numbersep=3pt,                  
    showspaces=false,                
    showstringspaces=false,
    showtabs=false,                  
    tabsize=2,
    moredelim=**[is][\color{red}]{&}{&},
    moredelim=**[is][\color{myblue}]{/}{/}
}
\usepackage{xcolor}
\definecolor{codegray}{rgb}{0.5,0.5,0.5}
\definecolor{codegreen}{rgb}{0,0.6,0}
\definecolor{codeblue}{rgb}{0,0,1}

\lstdefinestyle{mystyle}{
    basicstyle=\ttfamily\footnotesize,
    commentstyle=\color{codegreen},
    keywordstyle=\color{codeblue},
    showstringspaces=false,
    breaklines=true,
    frame=single,
    captionpos=b,
    tabsize=2,
    numbers=none,
    literate={**}{{\textbf{}}{}}2  
}
\usepackage{listings}
\usepackage{xcolor} 

\lstdefinestyle{prettyverilog}{
  language=Verilog,
  basicstyle=\ttfamily\small,
  keywordstyle=\color{blue}\bfseries,
  identifierstyle=\color{black},
  commentstyle=\color{gray}\itshape,
  stringstyle=\color{orange},
  numbers=left,
  numberstyle=\tiny\color{gray},
  stepnumber=1,
  numbersep=5pt,
  backgroundcolor=\color{white},
  showspaces=false,
  showstringspaces=false,
  showtabs=false,
  frame=single,
  tabsize=2,
  captionpos=b,
  breaklines=true,
  breakatwhitespace=false,
  escapeinside={(*@}{@*)}
}
\usepackage{listings}
\usepackage{xcolor} 

\lstdefinelanguage{json}{
    morestring=[b]",
    morecomment=[l]{//},
    moredelim=[s][\color{black}]{:}{,},
    stringstyle=\color{brown},
    keywordstyle=\color{blue},
    commentstyle=\color{gray}\ttfamily,
    morekeywords={true,false,null},
}

\lstdefinestyle{prettyjson}{
    language=json,
    basicstyle=\ttfamily\small,
    keywordstyle=\color{blue}\bfseries,
    stringstyle=\color{brown},
    commentstyle=\color{gray}\itshape,
    numberstyle=\tiny\color{gray},
    numbers=left,
    stepnumber=1,
    numbersep=5pt,
    backgroundcolor=\color{white},
    showstringspaces=false,
    breaklines=true,
    frame=single,
    tabsize=2,
    columns=fixed,
    captionpos=b
}
\usepackage{etoolbox}

\def\BibTeX{{\rm B\kern-.05em{\sc i\kern-.025em b}\kern-.08em
    T\kern-.1667em\lower.7ex\hbox{E}\kern-.125emX}}
\begin{document}

\title{SV-LLM: An Agentic Approach for SoC Security Verification using Large Language Models}

\author{
\IEEEauthorblockN{
Dipayan Saha, Shams Tarek, Hasan Al Shaikh, Khan Thamid Hasan, Pavan Sai Nalluri, Md. Ajoad Hasan,\\ Nashmin Alam, Jingbo Zhou, Sujan Kumar Saha, Mark Tehranipoor, and Farimah Farahmandi
}
\IEEEauthorblockA{
\textit{Department of Electrical and Computer Engineering, University of Florida, Gainesville, FL, USA}\\
\{dsaha, shams.tarek, hasanalshaikh, khanthamidhasan, pavansai.nalluri, md.hasan, nashminalam, jingbozhou, sujansaha\}@ufl.edu,\\ \{tehranipoor, farimah\}@ece.ufl.edu}
\thanks{This work was supported in part by the U.S. National Science Foundation (NSF) CAREER Award under Grant 2339971.}
}
\maketitle
\begin{abstract}
Ensuring the security of complex system-on-chips (SoCs) designs is a critical imperative, yet traditional verification techniques struggle to keep pace due to significant challenges in automation, scalability, comprehensiveness, and adaptability. The advent of large language models (LLMs), with their remarkable capabilities in natural language understanding, code generation, and advanced reasoning, presents a new paradigm for tackling these issues. Moving beyond monolithic models, an agentic approach allows for the creation of multi-agent systems where specialized LLMs collaborate to solve complex problems more effectively. Recognizing this opportunity, we introduce \textit{SV-LLM}, a novel multi-agent assistant system designed to automate and enhance SoC security verification. By integrating specialized agents for tasks like verification question answering, security asset identification, threat modeling, test plan and property generation, vulnerability detection, and simulation-based bug validation, \textit{SV-LLM} streamlines the workflow. To optimize their performance in these diverse tasks, agents leverage different learning paradigms, such as in-context learning, fine-tuning, and retrieval-augmented generation (RAG). The system aims to reduce manual intervention, improve accuracy, and accelerate security analysis, supporting proactive identification and mitigation of risks early in the design cycle. We demonstrate its potential to transform hardware security practices through illustrative case studies and experiments that showcase its applicability and efficacy.

\end{abstract}


\section{Introduction}
System-on-chips (SoCs) have seen widespread adoption across diverse application domains, including consumer electronics, IoT devices, healthcare equipment, industrial systems, and autonomous control platforms. 
This extensive adoption is largely attributed to the versatility inherent in SoCs: the integration of most or all components of the essential computer system onto a single chip enables them to deliver high computing performance while maintaining a small size and low power consumption.
However, with the growing diversity and volume of applications, demands on SoC performance have correspondingly escalated, leading to significant increases in design complexity. 
Simultaneously, the pressure to reduce time-to-market has shortened the available time frame for SoC integrators and design houses to thoroughly verify system functionality, performance, and security.

Despite decades of research and industrial effort, the verification of SoC hardware has remained predominantly manual. This challenge is reflected in industry trends, where, since 2007, the number of verification engineers has increased three times faster than that of design engineers \cite{darbari2024verification}.
Nevertheless, even with approximately 80\% of the design cycle dedicated to verification, 60–70\% of hardware development projects continue to fall behind schedule \cite{darbari2024verification}.
The growing prominence of security verification necessitates its integration as a fundamental aspect of hardware verification. This imperative has been reinforced by recent real-world exploits targeting commercially deployed SoCs, such as Pacman \cite{ravichandran2022pacman}, Augury \cite{vicarte2022augury}, and GhostWrite \cite{thomas2024riscvuzz}, which have demonstrated the critical security gaps that can arise when security considerations are deferred or not adequately addressed during the verification process.

An examination of contemporary security verification methodologies reveals that, much like trends in functional verification, the field remains largely manual, effort-intensive, and painstaking. Formal verification approaches \cite{rajendran2016formal, subramanyan2014formal} continue to dominate static security verification strategies. These techniques require engineers to possess substantial expertise in translating high-level security requirements into formal language assertions that can be evaluated by formal verification tools. However, formal methods face challenges with respect to scalability when applied to complex SoC designs and are prone to false positives\cite{nahiyan2017hardware}. 
Other static verification approaches, such as Concolic testing \cite{lyu2020scalable, lyu2019automated} and static code analysis \cite{kibria2022rtl,al2023quardtropy}, similarly scale poorly with large, heterogeneous SoC designs. 
Meanwhile, emerging dynamic verification techniques, including fuzz testing \cite{gohil2024mabfuzz, hossain2023socfuzzer, azar2022fuzz,trippel2021fuzzing} and penetration testing \cite{al2023sharpen, al2024re}, offer the advantage of runtime monitoring and coverage of execution paths that may be difficult to analyze statically. 
However, these approaches also require the manual definition of cost, objective, or feedback functions to guide test generation and mutation processes.
The crafting of such functions to accurately represent device security requirements, mathematically or ontologically, remains a complex task, and standardized methodologies or best practices have not yet emerged within the research community.

Moreover, security verification poses a unique challenge that functional verification does not: The threat landscape is continuously evolving. 
As new attack vectors are discovered and publicized, verification methods must adapt and generalize to remain effective~\cite{tarek2023benchmarking}. 
Thus, automation and adaptability are critical attributes of any forward-looking security verification strategy.

If we analyze all existing verification approaches from a high level, all of them share a fundamental limitation: although hardware security requirements are typically articulated in natural language during the specification phase, current verification methods depend on manually translating these requirements into intermediate formal or mathematical representations. This translation process introduces significant overhead, demands high levels of expertise, and increases the likelihood of human error.
Large language models (LLMs), with their advanced capabilities in interpreting and reasoning over natural language, offer a promising paradigm shift for security verification \cite{saha2024llm}. 
By streamlining or completely eliminating the translation process, LLMs could greatly reduce manual effort, improve reliability, and shorten verification cycles. 
Reflecting this potential, recent studies have begun exploring the use of LLMs for hardware security verification tasks \cite{saha2024llm, kande2024security, ahmad2024hardware,meng2024nspg,secrtllm,socurellm,saha2025threatlens,bugwhisperer,paria2024navigating,akyash2024self,10691745,faruque2024trojanwhisper,latibari2024automated,10904479,11022798}.
However, current LLM-based works remain limited in scope and are often narrowly focused on specific threat models or verification stages. 
Comprehensive security verification demands far more: it encompasses threat modeling and security asset identification, dynamic testbench or test suite generation aimed at maximizing security coverage, and the formulation of directed assertions or properties targeting specific, security-critical corner cases. 
\begin{figure*}[!t]
\centering
%\captionsetup{justification=centering}
\includegraphics[scale=.86]{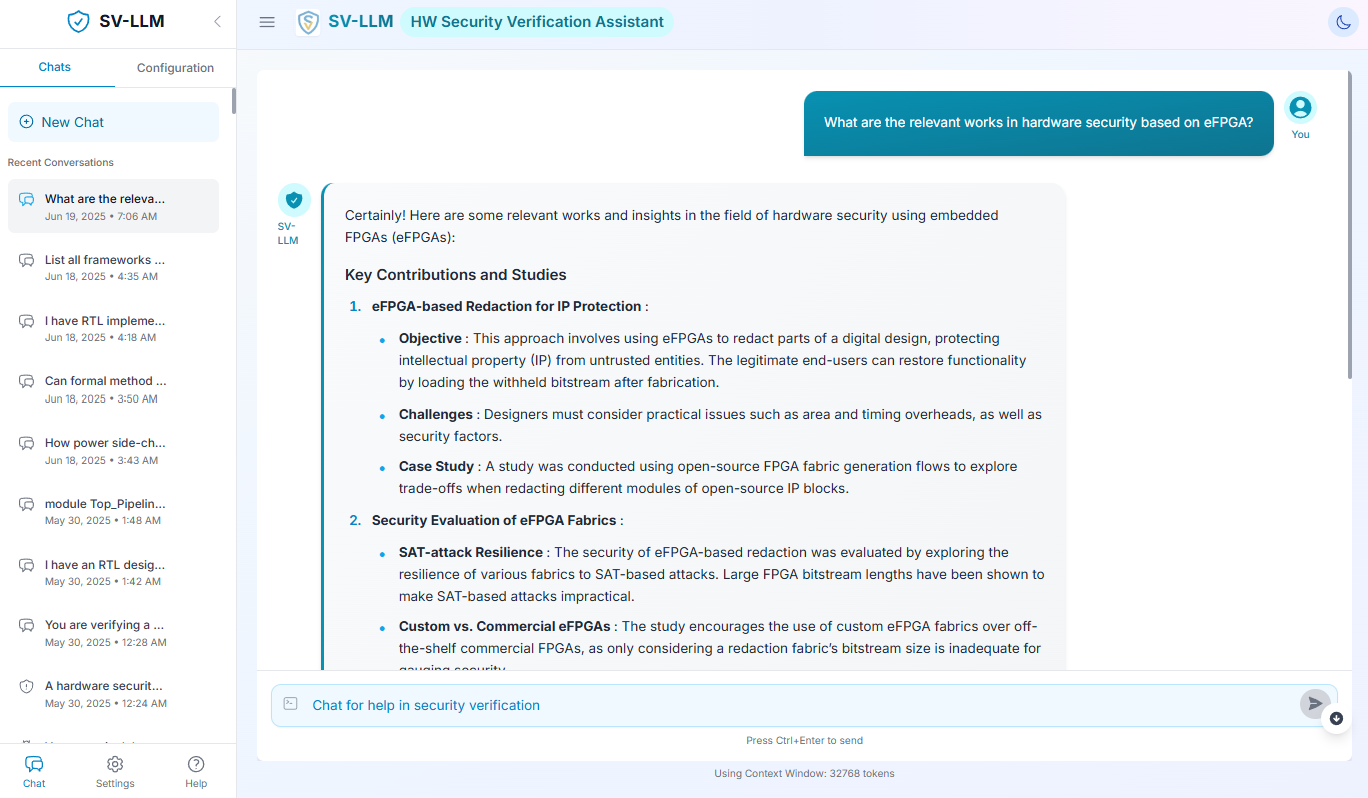}
\caption{Frontend interface of \textit{SV-LLM}.}
\label{fig:response_svllm}
%\vspace{-0.15in}
\end{figure*}
To fully realize the transformative potential of LLMs in hardware \textit{Security Verification}, in this paper, we introduce \textit{SV-LLM}, shown in Figure  \ref{fig:overview}, which comprehensively addresses this broader range of tasks in a systematic and scalable manner. 
\textit{SV-LLM} is a multi-agent framework designed to autonomously perform six key security verification tasks: security asset identification, threat modeling, test plan generation, vulnerability detection, security bug validation, and the generation of SystemVerilog properties and assertions for security verification.
In addition to automation, \textit{SV-LLM} features an interactive front-end chatbot interface that assists verification engineers in addressing complex security verification challenges, informed by current research developments from academia and industry. Unlike previous efforts to apply LLMs to hardware security, \textit{SV-LLM} distinguishes itself through its vastly broader applicability, dynamic adaptability, and holistic top-down approach to the verification workflow.

\textit{SV-LLM} is designed to support engineers with key aspects of the hardware security verification process, focusing on the register-transfer level (RTL). It assists in foundational tasks from security question answering to vulnerability analysis, while enabling continuous knowledge enrichment to address the challenges of the rapidly evolving hardware security landscape. A look of the frontend interface is shown in Figure \ref{fig:response_svllm}.

\section{Background}
As introduced in Section I, \textit{SV-LLM} harnesses the advanced capabilities of multi-agent LLM systems to execute a broad spectrum of verification tasks. This section provides foundational background on the concept of multi-agent LLM systems, outlines how they substantially extend the capabilities of individual LLMs, and explains the rationale behind adopting this paradigm within the \textit{SV-LLM} framework.

\subsection{Multi-agent LLM systems}
The agentic approach in artificial intelligence refers to designing systems that operate autonomously by perceiving their environment, reasoning over tasks, and executing actions toward predefined goals. Traditionally, these agents relied on symbolic planning, rule-based systems, or task-specific learning methods that lacked flexibility and generalization\cite{qiu2024llm}. 
The rise of LLMs has transformed this landscape by enabling a new class of agentic systems, in which a pre-trained LLM functions as the central reasoning core. These systems augment the LLM with external tools, long-term memory, and structured decision-making capabilities to support multi-step planning and goal-driven behavior. 
This transition has allowed agentic systems to handle more complex, open-ended problems, something earlier agents struggled with due to their limited scope and inability to adapt dynamically\cite{acharya2025agentic}.
Although LLM-driven single agent-based systems have shown considerable success in diverse domains, multi-agent systems have demonstrated even greater capability in handling extremely complex reasoning problems such as solving mathematical equations\cite{lei2024macm}, in defense against cyberattacks\cite{zeng2024autodefense}, financial decision making\cite{yu2024fincon}, software development\cite{manish2024autonomous, chen2024survey} etc.
This stems from the emphasis on diverse agent profiles and inter-agent communication, enabling a team of specialized agents to collaboratively address different sub-tasks.
By distributing tasks among cooperative LLM agents, these systems leverage domain-specific reasoning more effectively, especially in large-scale systems.

\subsection{Rationale for adoption in SV-LLM} 
In the context of hardware design and hardware security, multi-agent LLM frameworks provide clear advantages over single-prompt or monolithic agentic models. 
Hardware security tasks, such as threat modeling, security verification, or vulnerability detection, require deep contextual understanding, modular reasoning, and iterative refinement.
In addition, several verification tasks inherently depend on the outcomes of earlier stages of the verification workflow. For example, the generation of a security test plan is based on prior threat modeling, which itself is contingent upon the accurate identification of security assets. By adopting a multi-agent architecture, \textit{SV-LLM} enables the decomposition of complex verification tasks into manageable subgoals, supports collaborative decision-making among specialized agents responsible for individual subtasks, and allows agents to iteratively refine their outputs based on feedback - either from the verification engineer or from other agents - within an evolving verification context.

\section{SV-LLM}
\begin{figure*}[!t]
\centering
%\captionsetup{justification=centering}
\includegraphics[scale=.17]{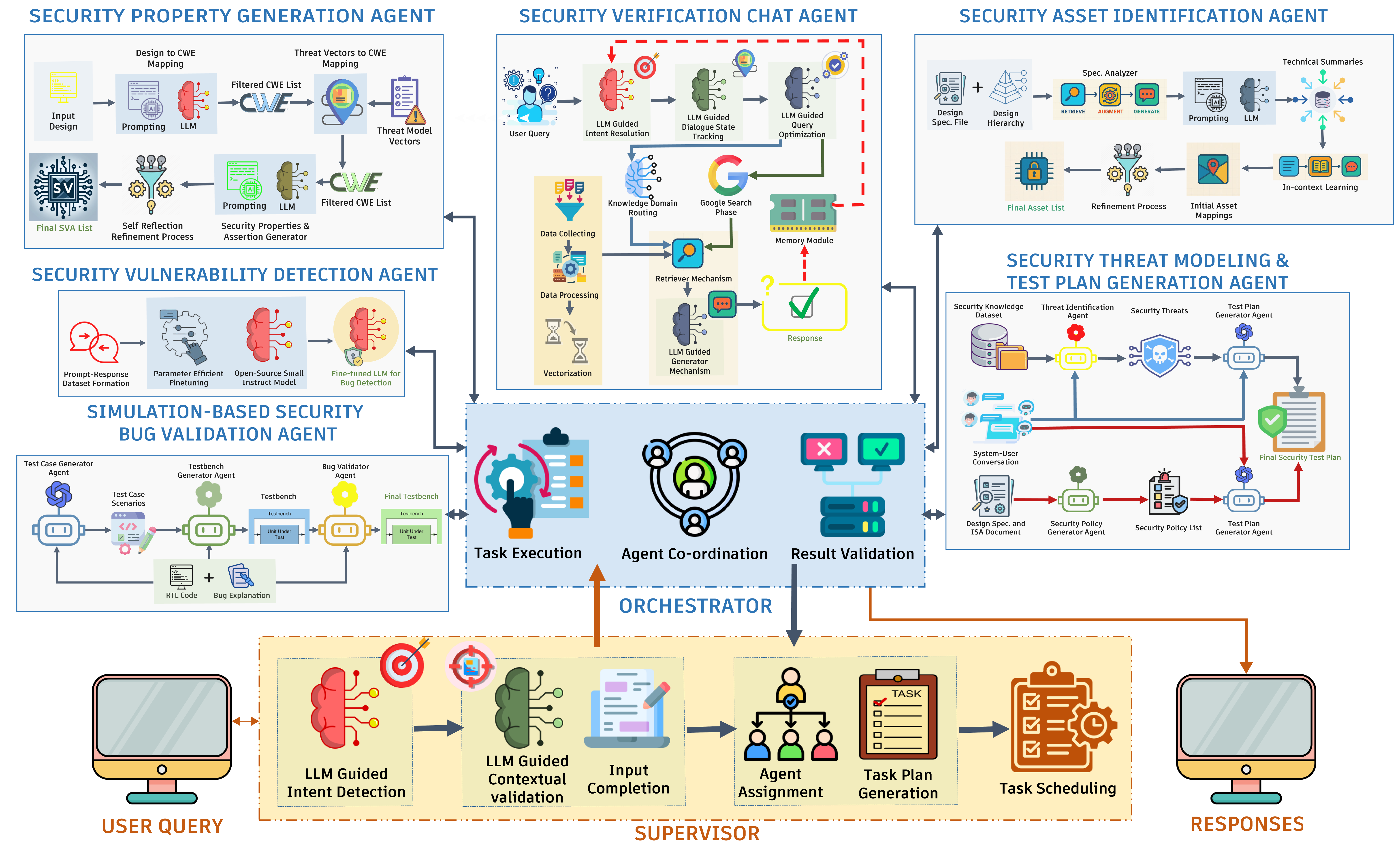}
\caption{Overview of \textit{SV-LLM}.}
\label{fig:overview}
%\vspace{-0.15in}
\end{figure*}

\textit{SV-LLM} is an LLM-driven agentic framework designed for automated security verification of hardware designs. It focuses on assisting verification engineers with several security verification tasks through natural language interaction. An overview of the methodology is shown in Figure \ref{fig:overview}.

\subsection{Key Features:}

\textit{SV-LLM} offers a suite of features designed to streamline and enhance the hardware security verification workflow through an LLM-driven agentic framework. The key features of the framework are outlined below.

\paragraph{Task-Oriented Agent Architecture}
At the core of \textit{SV-LLM} lies a task-centric architecture that supports a range of verification activities commonly encountered by hardware security engineers. The system is equipped with dedicated agents, each responsible for a specific class of tasks. These include:

\begin{itemize}
\item Security Q\&A: \textit{SV-LLM} can answer conceptual or practical questions about hardware security, verification methods, security threats, and other security-related topics.
\item Security Asset Identification: \textit{SV-LLM} is capable of analyzing the specification of an SoC design and recognizing critical security assets.
\item Security Property Generation: \textit{SV-LLM} is capable of generating formal security properties and corresponding assertions in SystemVerilog Assertion (SVA) format.
\item Threat Modeling and Test Plan Generation: Given the specification of an SoC design and the response of the user to certain queries, \textit{SV-LLM} can develop threat models and test strategies based on the features of a design and the attack surface.
\item Vulnerability Detection: \textit{SV-LLM} can analyze hardware designs and identify specific security vulnerabilities and weaknesses such as privilege escalation paths, insecure state transitions, etc.
\item Testbench Generation for Bug Validation: Given an RTL design and a bug identification report, \textit{SV-LLM} can construct simulation-based testbenches to validate the presence of specific security bugs.
\end{itemize}

\paragraph{Structured Output Format}
Each verification task produces results in a standardized and structured format, making the outputs easily usable in downstream tools or documentation pipelines. For example, generated properties are output in syntactically correct .sva files, suitable for direct inclusion in formal verification flows. Asset identification results are formatted as JSON objects, allowing for further analysis or integration into security review documentation. 
\begin{figure*}[t]
\centering
%\captionsetup{justification=centering}
\includegraphics[scale=.5]{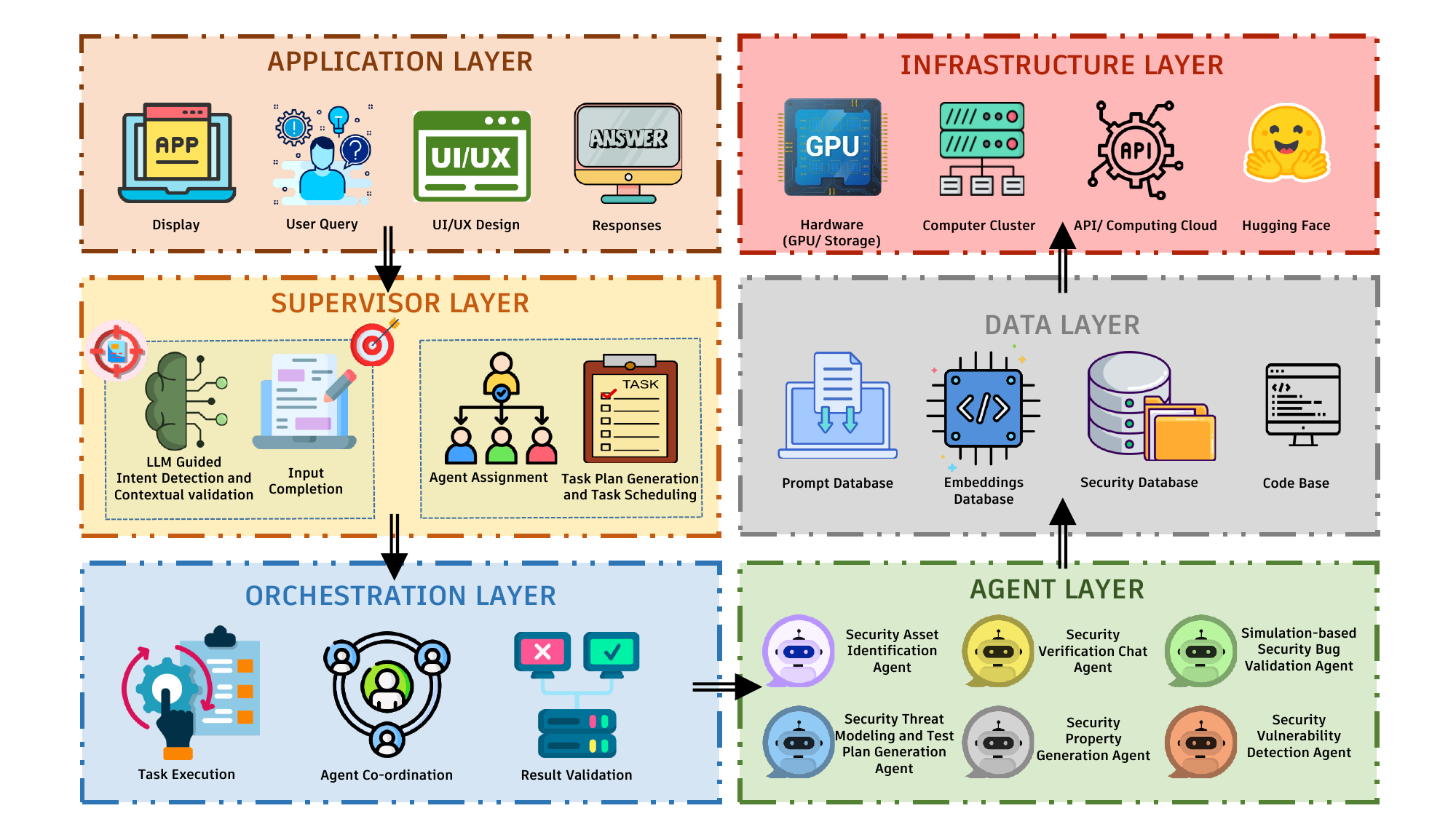}
\caption{Layered architecture of the \textit{SV-LLM} framework}
\label{fig:layers}
\end{figure*}
\paragraph{Continuous Dialogue and Iterative Refinement}
\textit{SV-LLM} supports a multi-turn dialogue structure, allowing users to refine their queries or follow up on previous results. After completion of a task, the user can immediately request further analysis, ask for explanations, or proceed to subsequent verification steps, without losing context. 

\paragraph{Safe and Domain-Constrained Dialogue}
To ensure the integrity and relevance of interactions, \textit{SV-LLM} actively constrains dialogue to topics within the scope of semiconductor design, hardware security, and system-level verification. Irrelevant or off-topic queries, particularly those unrelated to hardware, VLSI, or formal methods, are gracefully rejected, maintaining the professional focus of the tool.

\paragraph{Memory}
\textit{SV-LLM} incorporates both short-term and long-term memory mechanisms to preserve dialogue context across queries. While short-term memory allows for seamless continuity within a session, retaining user input, prior output, and the current state of the verification task, long-term memory stores the context of the overall conversation of the current session.
\subsection{Architecture of SV-LLM}
\begin{figure*}[t]
\centering
\includegraphics[scale=.11]{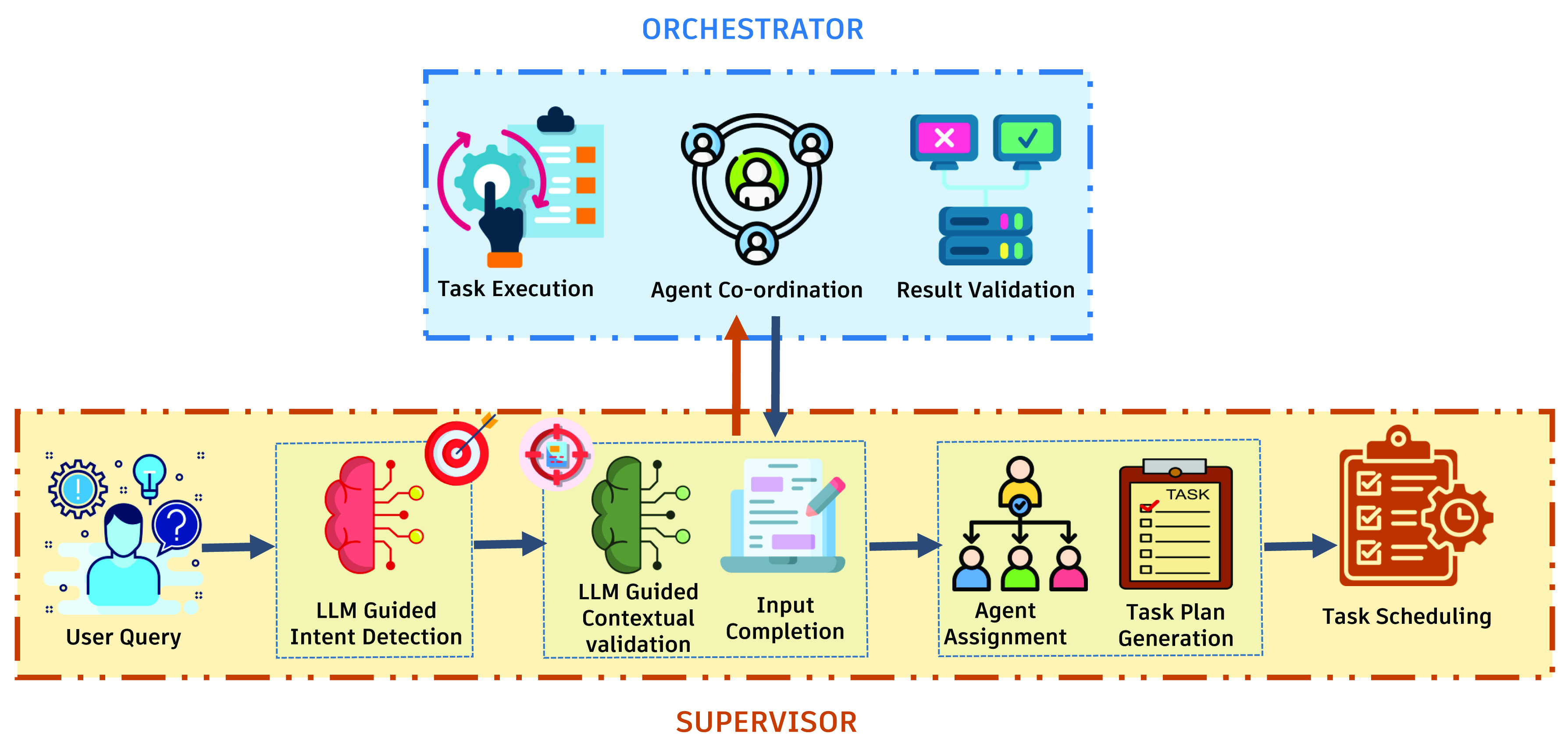}
\caption{Overview of Supervisor and Orchestrator.}
\label{fig:supervisor_and_orch}
\end{figure*}

Architecturally, SV-LLM is organized into six layered components shown in Figure \ref{fig:layers}. The Application Layer facilitates user interaction through a conversational and file-based interface, while the Supervisor Layer interprets user intent, completes missing context, and generates executable task plans. The Orchestrator Layer, shown in Figure \ref{fig:supervisor_and_orch}, ensures coordinated execution and validation across agents. At the core, the Agent Layer houses specialized modules for different tasks. Supporting these are the Data Layer, which maintains knowledge bases, design repositories, and embedding stores, and the Infrastructure Layer, which provides the computational backbone through APIs, GPU clusters, and hosted language models. The three main layers of the framework are described in the following.

\subsubsection{Supervisor Layer} 
At the core of the \textit{SV-LLM} framework lies the Supervisor, a dedicated control module responsible for interpreting user queries, validating contextual completeness, and preparing structured tasks for execution. Its primary aim is to bridge the gap between unstructured natural language input and modular, agent-driven verification workflows. By serving as the first point of processing, the Supervisor ensures that each user request is appropriately understood, fully contextualized, and systematically translated into a sequence of actions that can be executed by specialized agents.

The Supervisor plays a critical role in maintaining the robustness, modularity, and scalability of the \textit{SV-LLM} system. It enables flexible user interaction without compromising the formal rigor required for hardware security verification tasks. Moreover, by separating task preparation from task execution, the Supervisor establishes a clear abstraction boundary that supports extensibility and efficient coordination among system components.

\paragraph{LLM-Guided Intent Detection}
The first stage in the Supervisor’s processing pipeline is responsible for interpreting the user’s natural language query and identifying the underlying objective of the request. This step is critical in enabling the \textit{SV-LLM} framework to support a diverse range of hardware security verification tasks through a unified interface.

To accomplish this, the system employs a large language model to classify the query into one or more functional categories. These categories represent different verification-related activities, such as analyzing design vulnerabilities, generating formal properties, identifying security-relevant components, or guiding the creation of test strategies. In addition, the system distinguishes between informational queries and task-oriented requests, allowing it to differentiate conceptual questions from those that require concrete analysis or synthesis.
The language model further examines the query to determine whether any hardware design artifacts are referenced or included, such as RTL code or state machine descriptions, as well as to identify any mentions of known security vulnerabilities. This contextual understanding enables the Supervisor to assess the completeness of the request and to determine which specialized agent should be invoked to handle the task. Overall, this stage ensures that user queries are accurately interpreted and appropriately routed, forming the foundation for modular, task-specific processing within the broader \textit{SV-LLM} framework.

\paragraph{Contextual Validation and Input Completion}
Once the intent of the user query is established, the Supervisor proceeds to verify whether all information required to fulfill the associated task is available. This step is motivated by the observation that natural language queries are often incomplete, especially when users omit essential design artifacts or contextual constraints that are critical for executing specialized verification tasks.

The goal of this stage is twofold: first, to ensure that the input context is semantically complete and structurally adequate for downstream analysis; and second, to maintain the continuity of the interaction without forcing the user to rephrase or restart their query. This is particularly important in complex verification workflows where tasks such as vulnerability detection or property generation depend on access to well-formed hardware design descriptions, specifications, or threat-related inputs.

To achieve this, the Supervisor evaluates the query against a set of task-specific data requirements. These requirements reflect the minimal set of inputs needed for the task to proceed with meaningful results. If any required elements are missing or ambiguous, the system triggers a guided refinement loop, prompting the user to supply the missing components through an interactive interface. This dynamic querying mechanism ensures that task preparation remains user-friendly while preserving the integrity of automated verification. Through this validation process, the Supervisor ensures that only well-grounded and fully contextualized queries are forwarded for execution, thereby improving the reliability, interpretability, and effectiveness of the \textit{SV-LLM} framework.

\paragraph{Agent Assignment and Task Plan Generation}
Following intent detection and contextual validation, the Supervisor proceeds to assign the task to an appropriate agent within the \textit{SV-LLM} framework. Each agent is specialized to handle a particular class of hardware security verification tasks, and the assignment is determined directly by the intent previously inferred from the user query.

The purpose of agent assignment is to delegate execution to a modular component that is equipped with domain-specific reasoning aligned with the query's objective. This modularity supports task specialization and improves the maintainability and scalability of the overall framework. Once the appropriate agent is selected, the Supervisor generates a corresponding task plan. The task plan serves as an execution blueprint that guides the agent’s operation. It outlines a coherent set of steps tailored to the expected behavior of the agent and the nature of the task at hand. These steps reflect the logical decomposition of the user request into manageable actions that the agent can carry out reliably.

In addition to generating the task plan, the Supervisor is also responsible for scheduling its execution. This involves organizing the order of task initiation and coordinating with the Orchestrator to initiate runtime operations. Through this structured delegation process, the Supervisor ensures that user queries are transformed into actionable workflows, enabling the \textit{SV-LLM} framework to perform automated, goal-driven security verification.

\subsubsection{Orchestrator Layer}
Once the Supervisor has completed its responsibilities, including intent detection, input validation, agent assignment, and task plan generation, the Orchestrator assumes control of task execution. Serving as the runtime coordinator of the \textit{SV-LLM} framework, the Orchestrator is responsible for managing the operational flow of verification tasks and ensuring that each component in the execution pipeline functions in a coherent and efficient manner. The core function of the Orchestrator is to carry out the task plan produced by the Supervisor. This involves invoking the assigned agent and delegating individual sub-tasks to specialized sub-agents when applicable. Each sub-agent is designed to handle a specific operational unit within the broader task, enabling fine-grained modularity and parallelization where appropriate. The Orchestrator oversees this delegation process, tracks the execution state of each sub-task, and ensures that outputs are correctly routed between components.

In workflows involving multiple dependent steps or conditional branching, the Orchestrator maintains the execution logic and handles intermediate decisions based on agent outputs. It also plays a key role in managing run-time feedback: If a sub-task fails, the Orchestrator initiates corrective action. By decoupling execution from interpretation and planning, the Orchestrator introduces a clean separation of concerns that enhances system modularity, simplifies maintenance, and improves scalability. This design allows the \textit{SV-LLM} framework to accommodate increasingly complex verification workflows and integrate additional agents seamlessly in future expansions.

\subsubsection{Agent Layer} 
The Agent Layer serves as the computational backbone of \textit{SV-LLM}, composed of six specialized primary agents that collectively support the full spectrum of hardware security verification tasks. Each agent has been carefully designed with the complexity of its assigned task in mind, and leverages an appropriate learning paradigm—ranging from in-context learning for lightweight reasoning tasks, to fine-tuned models for vulnerability detection, and retrieval-augmented generation (RAG) for knowledge-intensive interactions. The six core agents include: \textit{Security Verification Chat Agent}, \textit{Security Asset Identification Agent}, \textit{Threat Modeling and Test Plan Generation Agent}, \textit{Vulnerability Detection Agent}, \textit{Simulation-Based Bug Validation Agent}, and \textit{Security Property Generation Agent}. Notably, each agent is composed of multiple sub-agents, each responsible for a finer-grained task. For example, the \textit{Threat Modeling and Test Plan Agent} includes a Threat Identification sub-agent, a Security Policy Generator, and a Test Plan Generator, all working in sequence. In addition to internal reasoning, several agents are designed to interface with external tools to complete their tasks—such as invoking a SystemVerilog syntax checker to validate assertions, using a ModelSim simulator to verify RTL behavior, or accessing a search engine to retrieve additional threat intelligence. This layered and task-aware agent design allows SV-LLM to perform robust, scalable, and intelligent security verification across diverse SoC design scenarios.

\section{Agents in SV-LLM}
\subsection{Security Verification Chat Agent} 
\begin{figure}[t]
\centering
%\captionsetup{justification=centering}
\includegraphics[scale=.4]{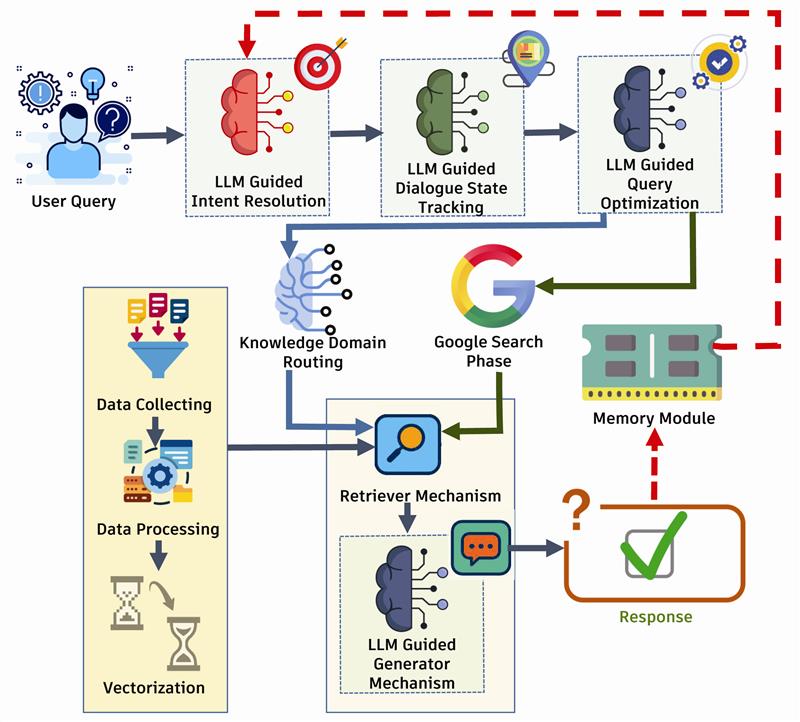}
\caption{Overview of Security Chat Agent.}
\label{fig:sv_chat}
\end{figure}

The \textit{Security Verification Chat Agent} is a modular, LLM-driven security verification chatbot designed to assist engineers and researchers in navigating complex hardware security challenges through natural language interaction. Figure \ref{fig:sv_chat} illustrates the overall system architecture of the agent, which comprises multiple interlinked components organized into three main stages: query understanding, information retrieval, and response generation.

Upon receiving a user query, the \textit{Security Verification Chat Agent} initiates the process with an LLM-guided intent resolution module. This module classifies the query into one of three categories: (i) security-related question, (ii) feedback to a prior response, or (iii) invalid/unsupported intent. If the intent is recognized as a valid query or feedback, the system proceeds to the dialogue state tracking module, which determines whether the query is a follow-up in a multi-turn interaction. This is achieved by linking the current query with previous turns using a memory module to maintain conversational coherence.

The next stage involves query optimization, where the input query is refined to improve retrieval quality. This includes syntactic simplification, entity normalization, and context-aware expansion using an LLM. The optimized query is then routed through two parallel information-gathering pathways. The first is a knowledge domain routing module, which classifies the query semantically and selects a relevant academic vectorstore constructed through offline data processing and vectorization. The second optional pathway leverages a Google search phase, enabling external knowledge injection when domain-specific information is insufficient or ambiguous. Both internal and external knowledge pathways feed into a retriever mechanism that filters and ranks relevant information. The retrieved content is then forwarded to the LLM-guided generator mechanism, which synthesizes a structured, contextually appropriate response. If the original query is a follow-up, the system also performs contextual reference resolution to anchor implicit references to earlier conversation history. The final response is then returned to the user, completing the dialogue turn.

The integration of offline domain knowledge, contextual memory, real-time search, and LLM-based natural language generation enables the \textit{Security Verification Chat Agent} to serve as a robust, intelligent assistant for security verification tasks in modern SoC and RTL design workflows.

\subsection{Security Asset Identification Agent}
\begin{figure*}[t]
\centering
%\captionsetup{justification=centering}
\includegraphics[scale=.1]{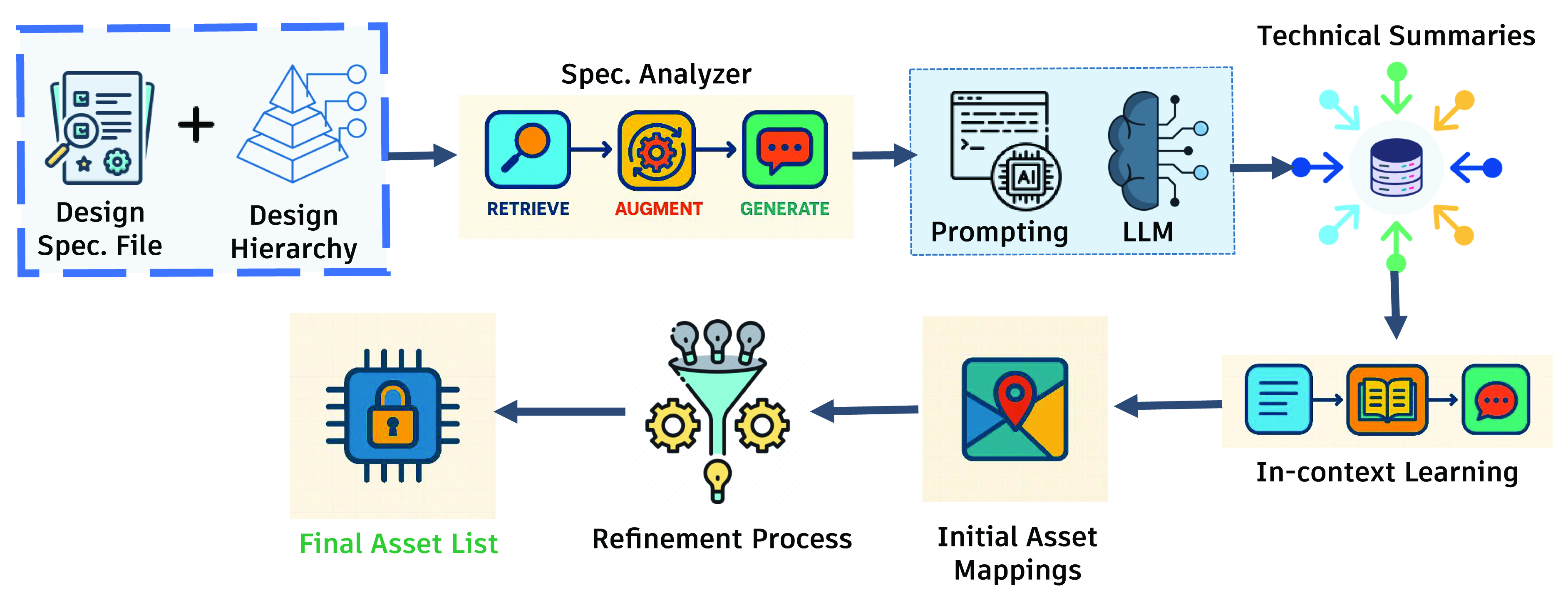}
\caption{Overview of \textit{Security Asset Generation Agent}.}
\label{fig:asset_gen}
\vspace{-0.15in}
\end{figure*}

The \textit{Security Asset Identification Agent}, shown in Figure \ref{fig:asset_gen}, streamlines the identification of security-critical components in the pre-silicon stage of SoC design. 
This agent is particularly crucial in \textit{SV-LLM} architecture, in the sense that early identification of assets facilitates more targeted threat modeling and security property generation.
Traditionally, asset identification relies heavily on individual expertise, making it error-prone and inconsistent. The need to automate this process has been emphasized in initiatives such as Accellera’s SA-EDI standard \cite{sa_edi} and the IEEE P3164 \cite{ieee_p3164} effort before. Existing approaches, such as \cite{nath2025toward} and \cite{ayalasomayajula2024automatic}, focus on asset identification from RTL alone, without utilizing the SoC specification. Although \cite{ayalasomayajula2024lasp} supplements RTL with specification files, all these methods fundamentally require RTL access, limiting their utility in scenarios where SoCs are delivered as grey boxes or RTL is incomplete. In contrast, we investigate fully automating potential asset identification only from the SoC hardware specification, making use of LLM's natural language processing capability.

The \textit{Security Asset Identification Agent} comprises of two key sub-agents:
\begin{itemize}
    \item Modular Spec Summarization: Here, we preprocess the spec file before initiating the ``\textit{Asset Generation}'' sub-agent. Otherwise, due to the limitation in token length and lethargy in memory retention while handling large specification files, ``\textit{Asset Generation}'' sub-agent would not go deep enough to analyze all the potential security-critical assets in a design, leading to erroneous results. The ``\textit{Modular Spec Summarization}'' sub-agent is built on RAG (Retrieve-Augment-Generate) based flow. The design modules are sequentially provided with a user query, whose response is used to augment the prompt for extracting the “\textit{Technical Summary}” using the GPT-4o model. This “\textit{Technical Summary}” consolidates all relevant information about a design module, enabling the Asset-LLM to determine whether it contains any security-critical assets.

    \item Asset Generation:
    Next, in \textit{Asset Generation} sub-agent, for each design module, we employ \textit{in-context learning} method to engineer the prompts in such a way that LLM can learn from the step-by-step case-specific scenario and examples about the CIA (Confidentiality, Integrity, and Availability) security objectives, then analyze the extracted \textit{Technical Summaries} to generate the asset information (if any) in a specified .json format. The response generated is again critiqued by another prompt and further revised so that false positive assets are excluded.
\end{itemize}

\subsection{Security Threat Modeling and Test plan Generation Agent}

\begin{figure}[t]
\centering
%\captionsetup{justification=centering}
\includegraphics[scale=.1]{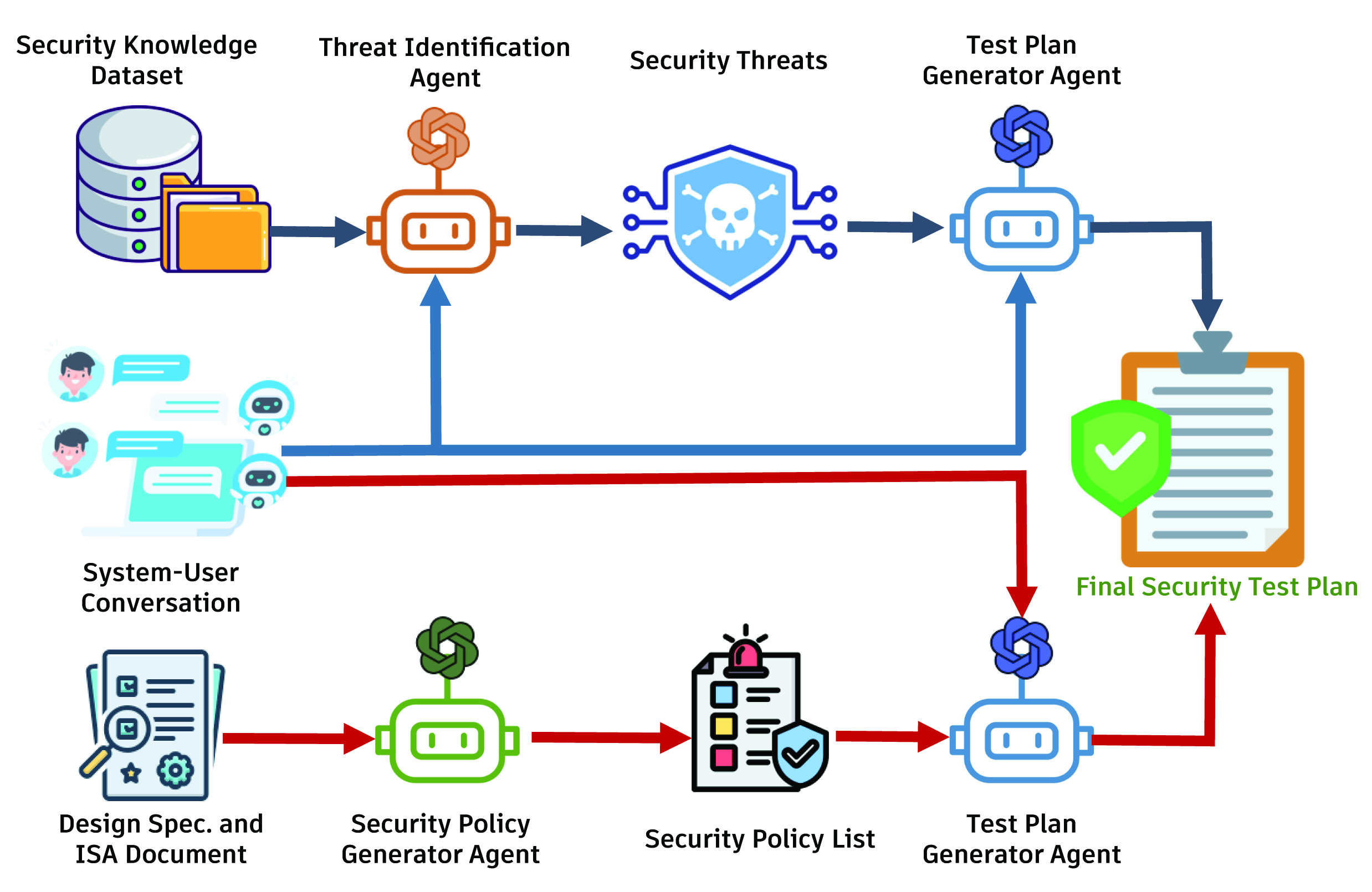}
\caption{Overview of Threat Modeling and Test Plan Generation Agent.}
\label{fig:threat_modeling}
\end{figure}

While the \textit{Security Asset Identification Agent} facilitates the identification of security-critical assets within the design, a comprehensive security verification effort also requires (i) the formulation of security requirements (i.e., policies) governing the flow and ownership of those assets, (ii) the modeling of threats that may compromise those requirements, and (iii) the construction of a detailed test plan capable of providing sufficient coverage against those threats. The \textit{Security Threat Modeling and Test Plan Generation Agent} in \textit{SV-LLM} is designed to achieve precisely these objectives. An overview of this agent is shown in Figure \ref{fig:threat_modeling}.
 
This agent starts with the design specification and the asset list provided by \textit{Security Asset Identification Agent} as inputs.
At a high level, it interprets these inputs and at the same time employs an LLM-based chatbot that collaborates interactively with verification engineers.
Through a series of iterative interactions and interpretation, a curated list of relevant threat models for the design is built.
Based on the threat models, it automatically generates structured security test plans.

At a more granular level, this agent employs a different workflow depending on the relevant threats to the design as shown in Figure \ref{fig:threat_modeling}.
These two workflows can be described as follows:
\begin{itemize}[leftmargin=*]
    \item Flow 1: This flow is engaged whenever the agent determines that the design is susceptible to physical and supply chain security threats such as side-channel attacks, laser fault injection attacks, clock glitching attacks, malicious design modifications, data remanence attacks, bus snooping,  hardware IP/IC cloning, counterfeit IC, reverse engineering, IC overproduction and other invasive and semi-invasive attacks. 
    This process begins with the \textit{Threat Identification} sub-agent, which retrieves relevant attack models from a curated knowledge base composed of academic publications and industry reports. 

    Following knowledge extraction, the sub-agent engages the verification engineer through a structured dialogue to collect system-specific details, including design characteristics, application context, and supply chain origins. 
    The sub-agent then evaluates the relevance of each potential threat based on the system’s context. 
    The sub-agent progresses this evaluation iteratively by repeatedly engaging the verification engineer. 
    The finalized threat list is then passed to another sub-agent called the \textit{Test Plan Generator}. The agent consults with the verification engineer to assess the available testing infrastructure, budget, and timelines, and subsequently formulates a security test plan. 
    \item Flow 2: The alternative flow of the agent addresses hardware vulnerabilities that are exploitable via software, such as privilege escalation, access control violations, and memory corruption. 
This flow is initiated by the \textit{Security Policy Generator} sub-agent, which extracts design-specific security policies from user-provided design specification documents and the assets identified by the upstream \textit{Security Asset Identifier} agent. 
Due to the length and complexity of these documents, the agent employs a RAG system to efficiently retrieve relevant content. 
First, registers, listed by the \textit{Security Asset Identifier} agent, are extracted from the specification using a retriever-LLM combination. 
Subsequently, a second RAG system searches the ISA document to extract corresponding security policies associated with each asset.
Once the policies are gathered, they are passed to the LLM of this sub-agent.  
It then separates the extracted policies, identifies their security significance, and highlights potential vulnerabilities related to them. The resulting policy list forms the basis for the subsequent agent.
\begin{figure}[t]
\centering
%\captionsetup{justification=centering}
\includegraphics[scale=.07]{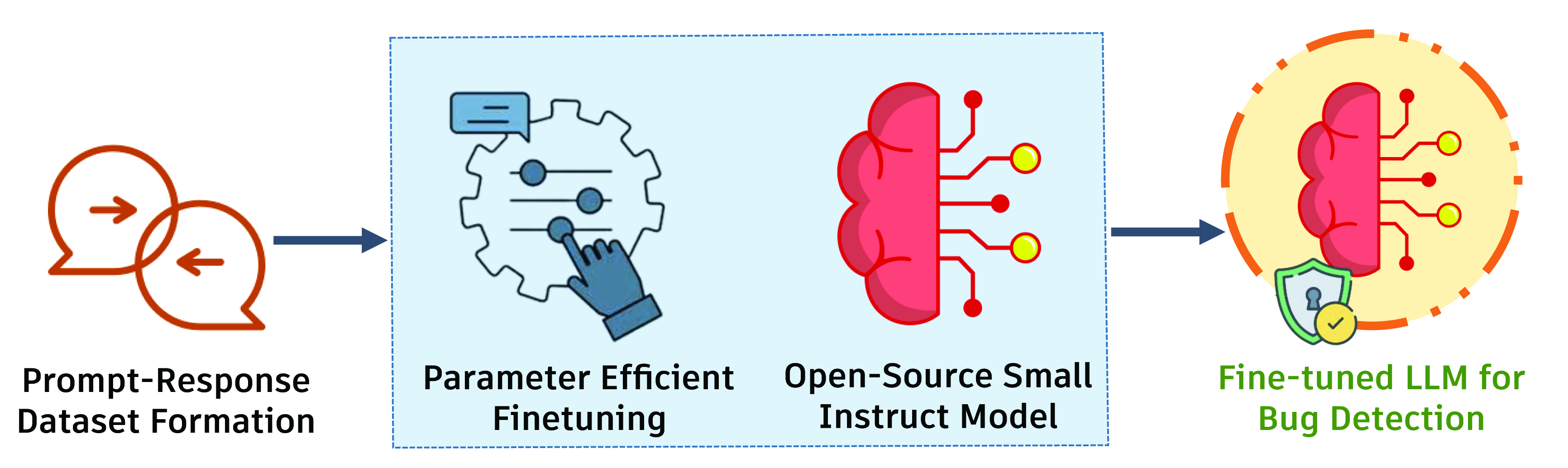}
\caption{Overview of \textit{Security Bug Detection Agent}.}
\label{fig:bug_detection_agent}
\end{figure}
\begin{figure*}[t]
\centering
%\captionsetup{justification=centering}
\includegraphics[scale=.1]{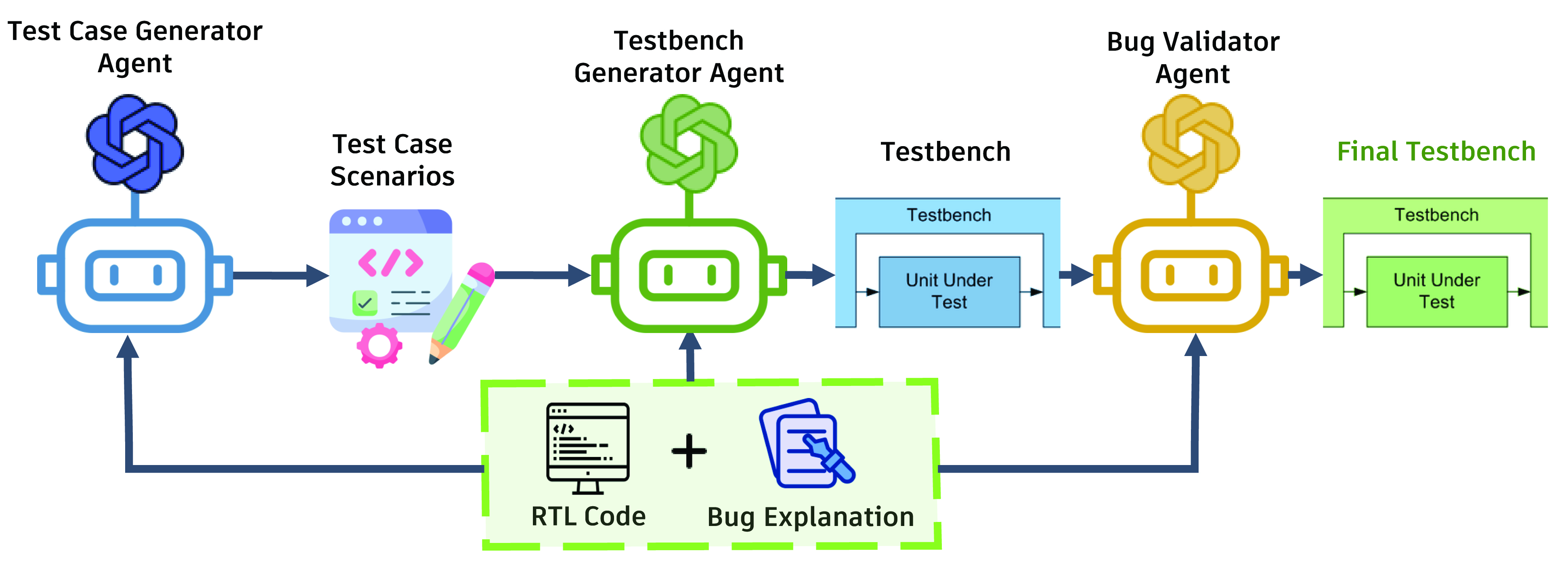}
\caption{Overview of \textit{Simulation-based Security Bug Validation Agent}.}
\label{fig:bug_validation}
\end{figure*}
The finalized security policies are then provided as input to the \textit{Test Plan Generator} sub-agent, which, consistent with its role in Flow 1, engages the verification engineer to assess available verification infrastructure and constraints. 
Using this information, the agent formulates a detailed security verification plan, specifying the targeted policy violations, verification objectives, methodologies, expected outcomes, and required tools.
\end{itemize}

Regardless of the flow undertaken, for each identified threat or policy, the agent generates detailed test cases that include:
objective of the test, methodology, expected behavior, evaluation criteria, tool recommendations

\subsection{Security Vulnerability Detection Agent}

The \textit{Security Vulnerability Detection Agent}, shown in Figure \ref{fig:bug_detection_agent}, is a core module of the \textit{SV-LLM} framework, specifically designed to automate the detection of security vulnerabilities in hardware designs at the RTL. Within the broader \textit{SV-LLM} ecosystem, which aims to enhance hardware security verification through a collection of specialized agents, the Bug Detection Agent addresses the critical challenge of identifying design-level security vulnerabilities efficiently and scalably. Traditional verification workflows are often manual and lack the adaptability needed for modern, complex SoC architectures. General-purpose LLMs, despite their general reasoning capabilities, fall short in domain-specific tasks such as vulnerability detection. To bridge this gap, the Bug Detection Agent leverages a custom fine-tuned open-source LLM, specifically adapted to understand and detect RTL security vulnerabilities.

As illustrated in Figure~\ref{fig:bug_detection_agent}, the foundation for the \textit{Security Vulnerability Detection Agent} was laid through a dedicated model preparation pipeline. First, a structured prompt-response dataset was constructed, where each example paired an RTL module with vulnerability-focused queries and annotated security evaluations. This dataset enabled parameter-efficient fine-tuning of an open-source small instruct model, ensuring that the model could internalize critical hardware security concepts while remaining lightweight and computationally efficient. The resulting fine-tuned LLM serves as the engine for the Bug Detection Agent, enabling it to perform targeted bug detection tasks with high accuracy and reliability.

During operation, the Bug Detection Agent receives RTL design inputs and formulates targeted security analysis queries based on known vulnerability patterns. These queries, together with the design context, are fed into the embedded fine-tuned LLM. The model then infers the presence or absence of specific vulnerabilities, providing detailed natural language explanations for its findings. This inference pipeline is entirely autonomous and does not involve any additional model fine-tuning at run-time. To maintain robustness, the agent applies context anchoring techniques and leverages model confidence estimation to filter uncertain or low-assurance outputs. Through this design, the Bug Detection Agent enables efficient, explainable, and scalable RTL security verification, thereby significantly contributing to \textit{SV-LLM}’s mission of democratizing automated hardware security analysis.

\subsection{Simulation-based Security Bug Validation Agent}
The \textit{Simulation-based Security Bug Validation Agent} is a vital component of our verification framework, specifically designed to confirm the presence of security bugs in RTL designs through automated testbench generation and simulation-based validation. Unlike generic testbench utilities, this agent treats testbench generation as a purposeful step in validating security exploitability, ensuring that each generated testbench meaningfully exercises the suspected vulnerability within a real simulation environment.

As shown in Figure \ref{fig:bug_validation}, the validation workflow is structured into three functional stages, coordinated by specialized subagents: the \textit{Test Scenario Generation Sub-agent}, the \textit{Testbench Generation Sub-agent}, and the \textit{Bug Validation Sub-agent}. 

The \textit{Test Scenario Generation Sub-agent} initiates the validation process by synthesizing temporally precise and semantically accurate scenarios tailored to activate the specified vulnerability. Using advanced capabilities of LLMs, the agent generates contextually coherent test events, detailing precise timing, relevant signal transitions, and explicit monitoring points necessary for effective vulnerability observation. An iterative feedback mechanism from an LLM-based critic further refines the generated scenarios, ensuring a comprehensive alignment with the vulnerability description and enhancing the likelihood of triggering the intended security flaw.

The \textit{Testbench Generation Sub-agent} subsequently transforms these validated scenarios into executable, simulation-ready SystemVerilog testbenches. This agent utilizes an iterative refinement loop driven by automated syntax checking and feedback integration, thereby ensuring both syntactic correctness and functional fidelity. By embedding essential monitoring constructs and validation logic directly into the generated testbenches, the agent facilitates precise observability and accurate detection of discrepancies in the expected hardware behavior during simulation.
\begin{figure*}[t]
\centering
%\captionsetup{justification=centering}
\includegraphics[scale=.1]{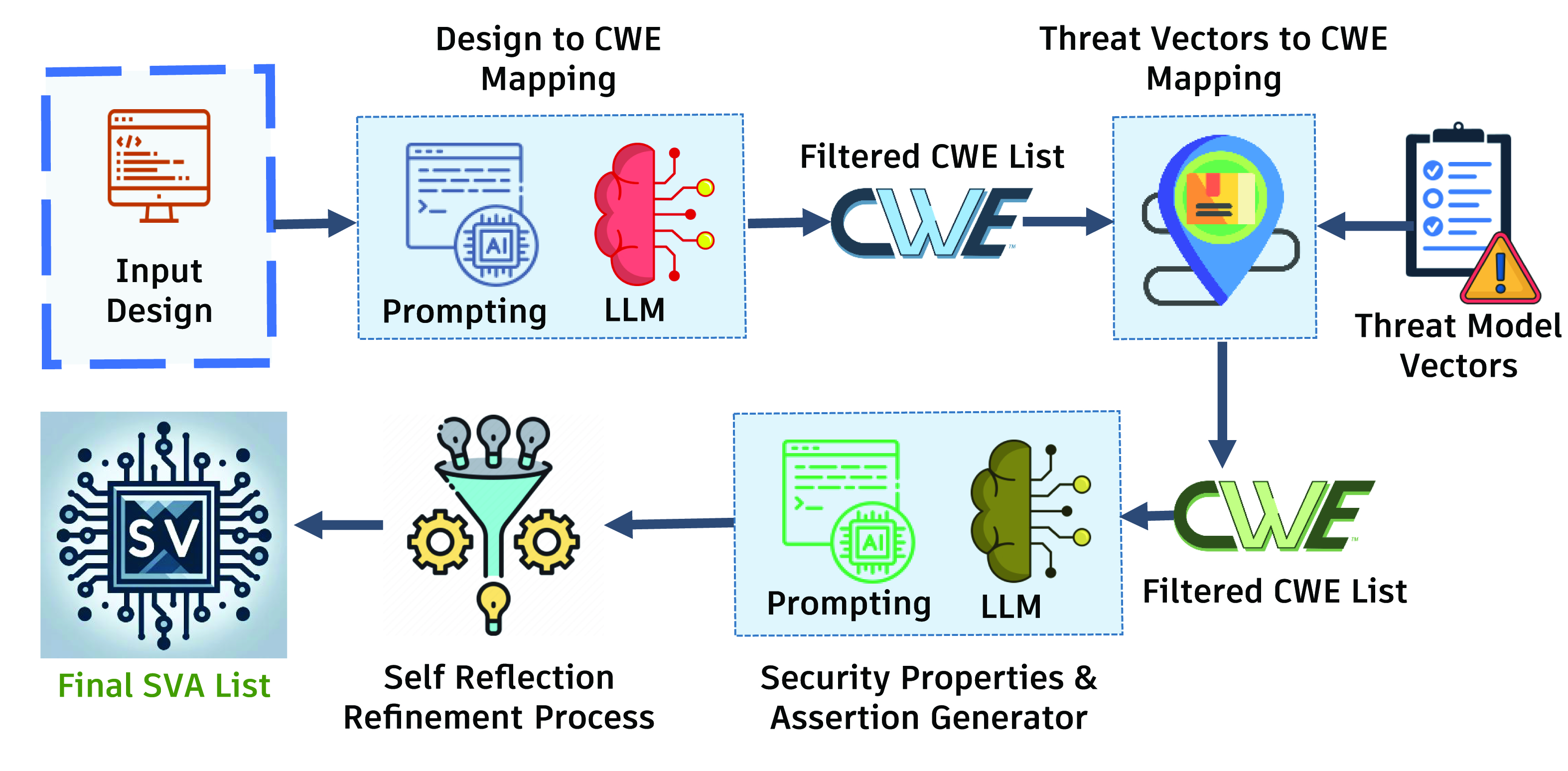}
\caption{Overview of \textit{Security Property Generation Agent}.}
\label{fig:property_gen}
%\vspace{-0.15in}
\end{figure*}

Finally, the {\textit{The Bug Validation Sub-agent} integrates simulation and analytical capabilities to conclusively verify the presence and correct manifestation of the targeted vulnerability. Through careful analysis of simulation results and comparison with predefined regions of interest (specific temporal and behavioral points critical for vulnerability verification), the sub-agent categorizes outcomes effectively into successful validation, failed activation, or incomplete definition scenarios. This robust validation mechanism ensures accurate and reliable identification of genuine vulnerabilities, significantly reducing false positives and negatives.

The output generated by the proposed agent framework includes precisely structured SystemVerilog testbenches that demonstrate essential characteristics such as syntactic correctness, functional readiness, and targeted observability. The testbenches are designed to explicitly activate and validate described security vulnerabilities, ensuring accurate runtime verification. Furthermore, comprehensive simulation results confirm the triggerability of vulnerabilities, providing detailed insight into the actual hardware behavior during execution and significantly enhancing the practical value of the generated validation artifacts. 

This agentic design enables our framework to directly connect the logical description of a bug to its runtime manifestation, closing the loop between LLM-driven testbench synthesis and simulation evidence. The agent operates without manual intervention, adapting to various types of bugs and design structures. Through this approach, the Security Bug Validation Agent ensures that each reported vulnerability is not only syntactically plausible but demonstrably observable in simulation, making the validation process both automated and trustworthy.

\subsection{Security Property and Assertion Generation Agent} 

The \textit{Security Property and Assertion Generation Agent} is a core component of the \textit{SV-LLM} framework, developed to automate the challenging and expertise-intensive task of formal security property generation. In conventional verification flows, writing formal properties requires deep domain expertise, significant manual effort, and extensive familiarity with both functional and security aspects of hardware designs. The challenge is even more pronounced for security property development, where correctness not only depends on design behavior but also on a precise understanding of potential threat models. Recent LLM-based approaches offer promise but often struggle with syntactic errors and context mismatches. To meet the demands of modern, fast-paced SoC design cycles, the Security Property and Assertion Generation Agent provides an automated, context-aware solution that generates syntactically correct, semantically valid, and tool-executable security properties and SVAs.
\begin{figure*}[t]
\centering
%\captionsetup{justification=centering}
\includegraphics[scale=.74]{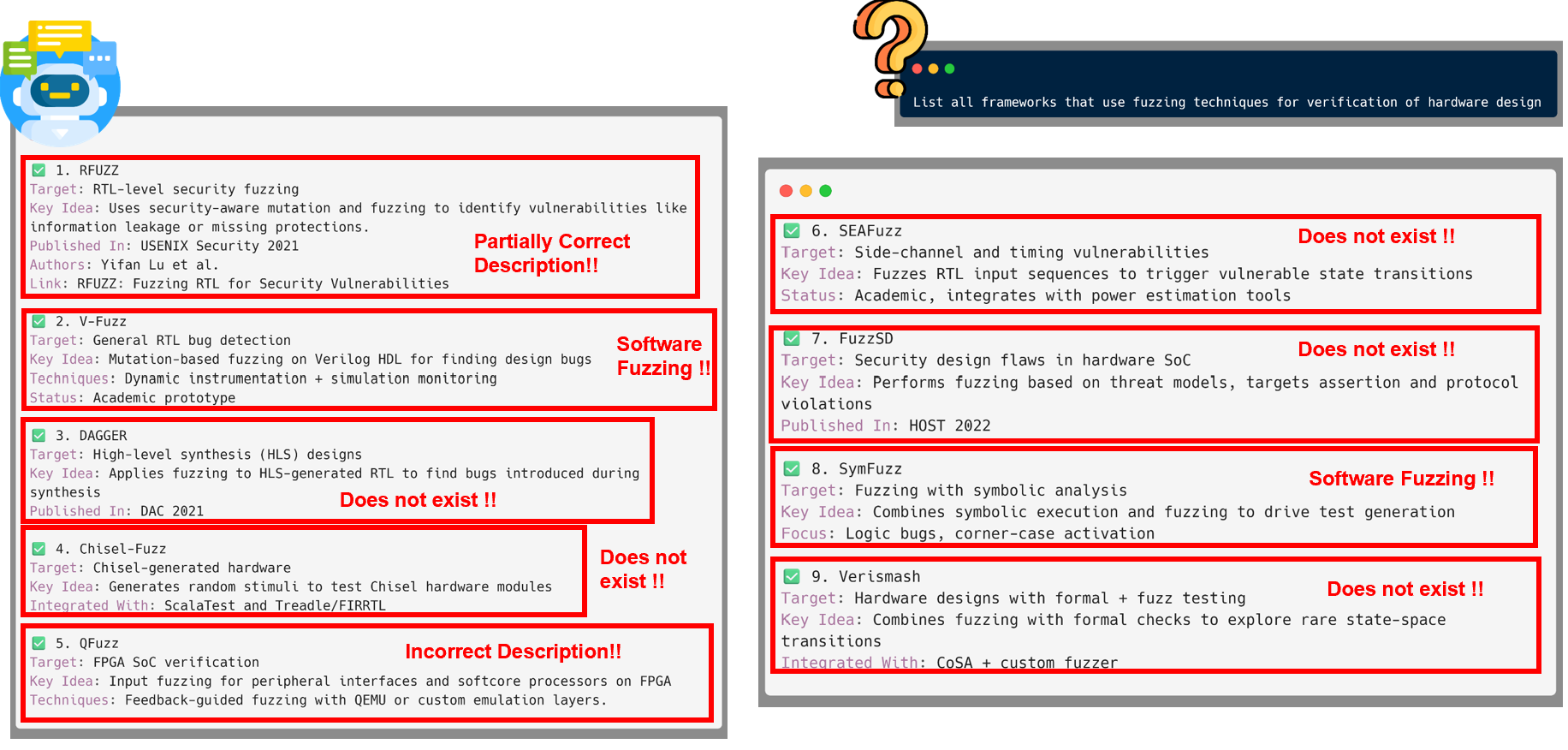}
\caption{Response of ChatGPT-4o.}
\label{fig:chatgpt_response}
\end{figure*}
\begin{figure*}[t]
\centering
%\captionsetup{justification=centering}
\includegraphics[scale=.5]{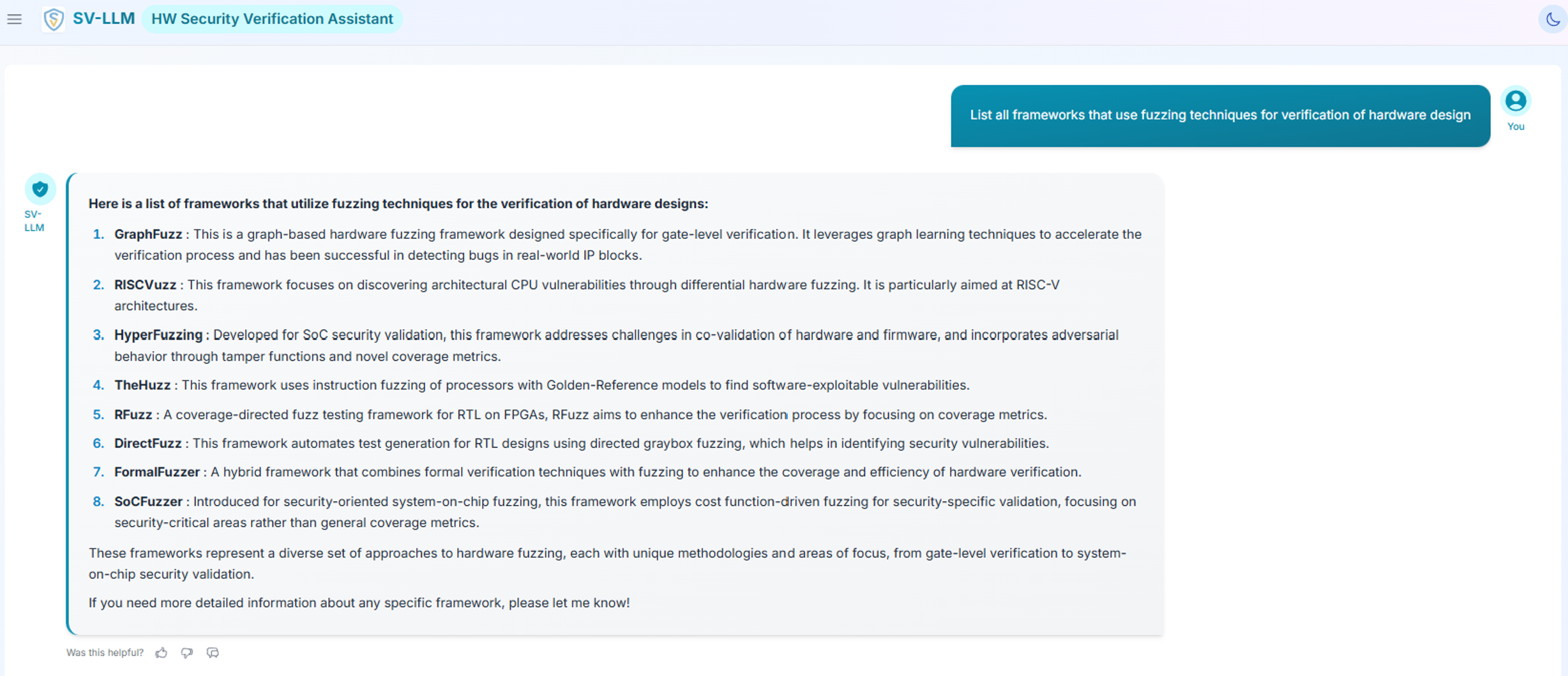}
\caption{Response of \textit{SV-LLM}.}
\label{fig:svllm_chat_response}
\end{figure*}
As depicted in Figure~\ref{fig:property_gen}, the agent operates through a structured multi-stage pipeline. The process begins with the Input Design being fed into a Design-to-CWE Mapping module. Here, a prompting engine and a specialized LLM classify the RTL design into predefined structural categories and associate it with relevant CWE identifiers. This produces a filtered CWE list that reflects the structural vulnerabilities of the design. Simultaneously, threat vectors provided by the user are processed through the Threat Vectors-to-CWE Mapping module. Using a static lookup approach grounded in domain knowledge and official CWE repositories, each threat vector is mapped to its potential CWE vulnerabilities. This yields a complementary filtered CWE list that captures the security threats applicable to the design.

The two filtered CWE lists, one derived from structural analysis and the other from threat modeling, are consolidated to form the input for the Security Properties and Assertion Generator. In this stage, another prompting-LLM engine crafts detailed prompts containing both signal-level design information and CWE-specific descriptions. The LLM then generates (i) detailed security scenarios demonstrating potential vulnerabilities, (ii) natural language security properties defining protection objectives, and (iii) executable SVA templates tailored to the design clocking and reset structures.

Finally, the outputs undergo a Self-Reflection Refinement Process to ensure formal correctness and practical executability. In this step, each generated SVA is automatically validated for syntactic soundness, signal consistency with the RTL design, and tool compatibility. Faulty, spurious, or misaligned properties are systematically filtered out. The final result is a validated high-quality SVA list, ready for direct use in formal verification workflows, significantly enhancing the scalability, correctness, and security rigor of the \textit{SV-LLM} framework.

\section{Case Studies}
To showcase SV-LLM's efficacy in performing the verification tasks described in earlier sections, we report case studies of each verification task in this section.
\subsection{Case Study I: Verification Q/A}

In a controlled comparison, we pose an identical question: ``List all frameworks that use fuzzing techniques for verification of hardware design" to two chatbots: ChatGPT-4o and \textit{SV-LLM}. The responses of these two chatbots are illustrated in Figure \ref{fig:chatgpt_response} and Figure \ref{fig:svllm_chat_response}.

The response of ChatGPT-4o, shown in Figure \ref{fig:chatgpt_response}, is not abstract or outdated; it is fundamentally incorrect and riddled with fabricated information. The model has produced a list of purported hardware fuzzing frameworks, the majority of which are complete hallucinations. For instance, frameworks it presented such as DAGGER, Chisel-Fuzz, SEAFuzz, FuzzSD, and Verismash do not actually exist. ChatGPT-4o had synthesized these names, randomly combining relevant-sounding concepts to create fictitious tools. In another instance of severe hallucination, while a tool named RFUZZ \cite{laeufer2018rfuzz} does exist, the model fabricated a portion of the associated details, including its publication venue, authors, and reference links. Furthermore, it misrepresented existing software verification tools such as V-Fuzz and SymFuzz, incorrectly presenting them as hardware fuzzing methodologies, demonstrating a critical lack of domain awareness. For a hardware security professional, such a response is actively harmful, leading to wasted time pursuing non-existent or irrelevant research.

In stark contrast, \textit{Security Verification Chat Agent} has delivered a response that is accurate, factually grounded, and directly aligned with the query. Guided by its RAG pipeline, the agent has retrieved relevant and verified information from its specialized knowledge base. It has provided detailed descriptions of established hardware-focused fuzzing frameworks such as \cite{laeufer2018rfuzz}, TheHuzz \cite{tyagi2022thehuzz}, SoCFuzzer \cite{hossain2023socfuzzer}, FormalFuzzer \cite{dipu2024formalfuzzer}, RISCVuzz \cite{thomas2024riscvuzz} and more. Each category and example is explicitly tied to published, peer-reviewed frameworks. Because its response is guided by retrieved information, \textit{SV-LLM} exhibited no hallucination. Beyond mere categorization, the chat agent's RAG architecture enabled it to inject precise, up‑to‑date details from its curated knowledge base. 

This case study underscores the severe limitations and potential dangers of using general-purpose chatbots for specialized hardware verification-related queries. Although ChatGPT-4o provided a veneer of confidence in its authority, its output is dangerously misleading. \textit{SV-LLM}, in contrast, has delivered depth, accuracy, and operational readiness, demonstrating that a RAG-based approach with a domain-centric knowledge base is essential to ensure the integrity of hardware security workflows.

\subsection{Case Study II: Identification of security assets}

As a case study for the Security Asset Identification Agent, we have chosen the NEORV32 32-bit RISC-V processor, an open-source SoC comprising a diverse set of core modules (e.g., CPU, memory, and debug modules), along with several security-critical peripheral modules such as TRNG, DMA, and interrupt controllers. The processor specification document provides detailed configurations and functional descriptions of various elements within each module, including signals, registers, and more.

\begin{figure*}
\centering
%\captionsetup{justification=centering}
%\includegraphics[scale=.54]
\lstset{style=prettyjson}
\begin{lstlisting}[caption={A sample of generated assets for NEORV 32 bit processor}, basicstyle=\ttfamily\scriptsize, label=lst:json]
{
    "IP": "cpu_cp_crypto",
    "Assets": [
        {
            "Asset_Name": "ShangMi Block Cipher Instructions",
            "Functionality": "Implements ShangMi block cipher instructions, executing in 6 cycles.",
            "Security Objective": "Confidentiality",
            "Justification": "The ShangMi block cipher instructions are used for encrypting data, which is crucial for maintaining the confidentiality of the data processed by the IP."
        }
    ]
}{
    "IP": "pwm",
    "Assets": [
        {
            "Asset_Name": "PWM_CFG_CDIV",
            "Functionality": "Divides a 10-bit clock for fine frequency tuning.",
            "Security Objective": "Integrity",
            "Justification": "The integrity of the PWM_CFG_CDIV field is critical because unauthorized changes could alter the PWM frequency, affecting the performance and behavior of the PWM-controlled devices."
        }
    ]
}{
    "IP": "cpu_decompressor",
    "Assets": [
        {
            "Asset_Name": "Decompression Logic",
            "Functionality": "Converts compressed 16-bit RISC-V instructions into their full 32-bit equivalents using the RISC-V 'C' extension.",
            "Security Objective": "Integrity",
            "Justification": "The decompression logic must accurately translate compressed instructions to ensure correct execution. Any modification could lead to incorrect instruction execution, affecting the processor's operation."
        }
    ]
}
\end{lstlisting}
\end{figure*}
As per the standard execution flow, the agent first extracted and refined the design  hierarchy, pruning modules like package (neorv32\_package), glue (neorv32\_top) or image (neorv32\_application\_image) modules. Then, the agent invoked RAG to derive separate `` Technical Summaries'' from the specification, for each NEORV32 module, i.e. Watchdog Timer, TRNG, UART etc .  For example, the TRNG summary would capture the textual description of registers/flags (\textit{TRNG\_CTRL\_EN}, \textit{TRNG\_CTRL\_FIFO\_MSB} etc.) configuration and their operational interaction at the modular and processor level.

Next, after being trained with in-context CIA examples of well-known hardware IPs (e.g., GPIO, AES, etc.), the agent proposed a list of candidate assets, which were further refined through self-critique prompts. The final output file for NEORV32, a sample of which is shown in Listing 1, listed key assets across security-relevant modules. For instance, the \textit{Decompression Logic}, \textit{Instruction Fetch Interface} and \textit{Instruction Dispatch Interface} were identified as assets in the Compressed Instructions Decoder module. 
Moreover, the \textit{Decompression Logic} was flagged as integrity-critical to translate compressed instructions accurately. A sample of assets generated for the NEORV32 is shown in Listing 1. This case study demonstrates the agent’s ability to autonomously identify security-critical assets across a complex SoC design using only its specification document and nothing else, which would greatly reduce the manual effort traditionally required for this purpose.

\subsection{Case Study III: Generation of Security Property and Assertion}

To evaluate the Security Property and Assertion Generation Agent, we applied it to a representative SoC subsystem (\textit{uart\_dma\_top}) integrating a UART module, a DMA controller, and a debug bridge. The design poses several security risks due to its plaintext UART echo, unrestricted debug access to critical configuration registers, and absence of privilege checks on memory-mapped register writes. These characteristics make it an ideal candidate for testing the agent’s ability to automatically generate semantically valid and security-aware SVAs.

Following the standard agent workflow, the design was classified as including both a DMA controller and a debug interface. The agent mapped the design to multiple relevant CWE classes, such as \textit{CWE-284 (Improper Access Control)} and \textit{CWE-1244 (Unlocking Debug Features Without Authorization)}. Simultaneously, the user-supplied threat vector, "Improper Access Control", was mapped to overlapping CWE identifiers. Upon intersecting the two lists, the agent generated a set of security scenarios, natural language security properties (NL-Properties), and fully executable SVAs tied to key design signals such as \textit{dbg\_sel}, \textit{dbg\_en}, \textit{dbg\_rdata}, and the \textit{csr\_q} configuration register.

The generated properties, shown in Listing 2, capture both confidentiality and access control requirements. For instance, the first property ensures that sensitive configuration values such as DMA enable and priority settings are cleared when debug mode is active. This helps detect bugs where the design may leak operational states or critical control values during a debug session. The second property enforces the masking of the debug output. These assertions serve as guards against unintentional information exposure through \textit{dbg\_rdata}, a common leakage channel. Additionally, the agent generated confirmatory properties ensuring that even legitimate debug interactions do not reveal sensitive data-e.g., assigning nonsensitive, fixed constants like \texttt{0xCAFEBABE} to confirm writes.

These security properties are particularly valuable for catching subtle information leakage bugs and access control violations that are often overlooked in traditional functional verification. By automatically generating properties grounded in security context and RTL signal semantics, the agent significantly reduces the manual effort and domain expertise required to build high-assurance security verification environments. This case study demonstrates the practical value of integrating the Security Property and Assertion Generation Agent into SoC verification flows to proactively secure hardware against privilege escalation and debug interface exploitation.
\begin{figure}
\centering
\lstset{caption={Generated Security Properties for the input UART Design}, label=sva} 
\begin{lstlisting} [style={prettyverilog}, linewidth=250pt]
    assert property (@(posedge clk) 
        disable iff (!rst_n) 
        (dbg_sel && dbg_en) |-> 
        (csr_q.enable_dma == 1'b0 && 
        csr_q.dma_prio == 3'h0));
    endproperty

    assert property @(posedge clk) 
        disable iff (!rst_n) 
        (dbg_sel && dbg_en) |-> 
        (dbg_rdata == 32'hDEADBEEF 
        || dbg_rdata == 32'hCAFEBABE);
    endproperty
    
    assert property (@(posedge clk) 
        (dbg_sel && dbg_en) |-> 
        (dbg_rdata == 32'hDEADBEEF));
    endproperty
    
    assert property (@(posedge clk) 
         (dbg_sel && dbg_en) |-> 
         (dbg_rdata == 32'hCAFEBABE));
    endproperty
\end{lstlisting}
\end{figure}
\subsection{Case Study IV: Detection of Security Vulnerability}

The first case study demonstrates the Bug Detection Agent's effectiveness in analyzing a password verification module, \textit{Authentication\_Bypass}, designed to authenticate users by comparing two 8-bit hash values. The module accepts a clock, reset, and validation signal as inputs, along with the input and correct hash values. It produces two outputs: \textit{authenticationFlag}, indicating a successful authentication if the hashes match, and \textit{nextState}, controlling finite state machine (FSM) transitions. The FSM comprises two main states: \textit{AuthenticateState}, where the module waits for a valid hash input, and \textit{WaitState}, which follows a verification attempt.

To evaluate the Bug Detection Agent, the RTL design was provided as input, as depicted in Figure \ref{fig:case_study_detection}. The agent analyzed the design’s behavior, particularly the FSM transitions based on hash validation. The agent identified a critical security vulnerability: irrespective of the correctness of the authentication result, the design transitions to the \textit{WaitState}. Even if the \textit{inputHash} does not match the \textit{correctHash}, the system progresses without terminating or blocking failed attempts, violating expected security practices. In a secure design, failure to authenticate should prevent further access attempts or trigger a secure reset.

\begin{figure}
\begin{tcolorbox}[colback=blue!10!white, colframe=gray!50!blue, title=\textbf{\textit{SV-LLM} Response for Hardware Vulnerability Detection}]
\label{vul_agent}
The code snippet provided is a part of a Password Verification Module, specifically designed to authenticate user credentials. However, it contains a critical flaw in its implementation of the authentication process, which could potentially lead to security vulnerabilities.
\newline The issue lies in the fact that the module transitions to the \textit{WaitState} regardless of whether the \textit{authenticationFlag} is set to \textit{1'b1} or \textit{1'b0}. This means that if the \textit{inputHash} does not match the \textit{correctHash}, the authentication process should ideally be terminated, and the user should be prompted to re-enter their credentials. However, the code continues to the \textit{WaitState}, which could be exploited by an attacker to bypass the authentication mechanism.
\newline In essence, the flaw allows an attacker to attempt multiple authentication attempts without being blocked, increasing the risk of unauthorized access. This is a significant security concern, as it could enable an attacker to gain access to sensitive resources or systems protected by this authentication mechanism.
\end{tcolorbox}
\caption{\textit{SV-LLM} Response for Hardware Vulnerability Detection}
\label{fig:case_study_detection}
\end{figure}

The Bug Detection Agent accurately pinpointed the flaw in natural language, explaining that this behavior enables an attacker to attempt multiple credential guesses without proper blocking. This security hole could be exploited to perform brute-force attacks on the authentication mechanism, significantly undermining the integrity of the system. The agent response not only identified the flaw but also contextualized it within a larger threat model, highlighting risks associated with improper FSM transition logic.

This case study highlights the ability of the bug detection agent to uncover subtle security flaws at the RTL level and generate human-readable explanations, thus facilitating rapid vulnerability detection and improving trust in automated security verification workflows.

\subsection{Case Study V: Validation of security bugs}

This case study evaluates the effectiveness of our automated security verification framework by analyzing an authentication finite state machine (FSM) module presented in Listing \ref{lst:rtl_fsm}. This RTL design is intended for user authentication via cryptographic hash validation, transitioning between states: IDLE, AUTHENTICATE, and WAIT\_STATE. However, it contains a critical vulnerability described in Figure \ref{fig:bug_description}, where the FSM erroneously transitions to WAIT STATE regardless of the authentication result, potentially allowing unauthorized access.
\begin{figure}[t]
\centering
\lstset{style=prettyverilog}
\begin{lstlisting}[caption={RTL Authentication Module with Bug in State Transition}, label=lst:rtl_fsm]
module Authentication Bypass (
    input clk,
    input rst_n,
    input isHashValid,
    input [127:0] inputHash,
    input [127:0] correctHash,
    output reg authenticationFlag
);
    parameter IDLE = 2'b00, AUTHENTICATE = 2'b01, 
    WAIT_STATE = 2'b10;
    reg [1:0] currentState, nextState;
    always @(posedge clk or negedge rst_n) begin
        if (!rst_n) begin
            currentState <= IDLE;
            nextState <= IDLE;
        end else begin
            currentState <= nextState;
        end
    end
    always @(*) begin
        nextState = currentState;
        authenticationFlag = 1'b0;
        case (currentState)
            IDLE: nextState = AUTHENTICATE;
            AUTHENTICATE: begin
                if (isHashValid) begin
                    if (inputHash == correctHash)
                        authenticationFlag = 1'b1;
                    else
                        authenticationFlag = 1'b0;
                    nextState = WAIT_STATE; 
                end else
                    nextState = AUTHENTICATE;
            end
            WAIT_STATE: nextState = IDLE;
            default: nextState = IDLE;
        endcase
    end
endmodule
\end{lstlisting}
\vspace{-0.2in}
\end{figure}
\begin{figure}[t]
\centering
\begin{tcolorbox}[colback=red!5!white, colframe=red!70!black, title=Security Flaw in FSM]
\scriptsize
In this revised code, the logic for authenticating a password is incorrectly implemented, leading to a critical flaw. Regardless of whether the \texttt{authenticationFlag} indicates a successful or failed password match, the system transitions to a waiting state. This flaw could allow unauthorized users to circumvent security measures by exploiting the incorrect handling of authentication states, posing a significant security risk.
\end{tcolorbox}
\caption{Description of the authentication FSM vulnerability}
\label{fig:bug_description}
\vspace{-.1in}
\end{figure} To systematically confirm this vulnerability, the \textit{Test Case Generator Agent} within this framework constructed detailed simulation scenarios, illustrated in Figure \ref{fig:test_scrnario}. Initially, the FSM was reset at 0 ns and the reset signal was released at 10 ns. At 20 ns, a valid authentication scenario transitioned the FSM from AUTHENTICATE to WAIT\_STATE with the authenticationFlag correctly asserted at 25 ns. Subsequently, at 35 ns, an invalid authentication scenario was introduced. Incorrect credentials were provided at 40 ns, but notably, the FSM incorrectly transitioned into WAIT\_STATE at 45 ns despite the authenticationFlag being properly deasserted, thus clearly illustrating the vulnerability.

Following scenario generation, the \textit{Testbench Generator Agent} of this framework produced an executable SystemVerilog testbench (Listing \ref{lst:tb}). The generated testbench included accurate clock generation, reset sequences, and input stimuli corresponding precisely to the defined scenarios. Comprehensive logging of FSM states, inputs, and outputs was achieved using \$strobe statements triggered at positive clock edges.
\begin{figure} 
\centering
\begin{tcolorbox}[colback=blue!5!white, colframe=blue!60!black, title=Step-by-Step Simulation Flow with Signal States]
\scriptsize
\begin{itemize}[leftmargin=*]
  \item \textbf{0 ns:} Assert reset. \texttt{clk=0, rst\_n=0, isHashValid=0}, all hash inputs cleared.
  \item \textbf{10 ns:} Release reset. \texttt{rst\_n=1}, waiting for clock edge.
  \item \textbf{15 ns:} First rising edge. Enters \texttt{IDLE}.
  \item \textbf{20 ns:} Apply valid credentials. \texttt{inputHash = correctHash, isHashValid = 1}.
  \item \textbf{25 ns:} Expect \texttt{authenticationFlag = 1}. FSM transitions to \texttt{WAIT\_STATE}.
  \item \textbf{30 ns:} Transition from \texttt{WAIT\_STATE} back to \texttt{IDLE}.
  \item \textbf{35 ns:} Prepare invalid auth case. Reset \texttt{isHashValid = 0}.
  \item \textbf{40 ns:} Assert incorrect \texttt{inputHash} with \texttt{isHashValid = 1}.
  \item \textbf{45 ns:} \textbf{BUG:} Although \texttt{authenticationFlag = 0}, FSM still enters \texttt{WAIT\_STATE}.
  \item \textbf{50 ns:} FSM returns to \texttt{IDLE}. Bug confirmed persistent.
  \item \textbf{55 ns:} Edge case: Random hash with \texttt{isHashValid = 0}. Monitor state behavior.
  \item \textbf{60 ns:} Final stability check: no transition, remains in \texttt{IDLE}.
\end{itemize}
\end{tcolorbox}
\caption{Test scenarios generated for the FSM bug, with precise timestamps and signal transitions to trigger and observe the flaw}
\label{fig:test_scrnario}
\end{figure}
\begin{figure}
\centering
\lstset{style=prettyverilog}
\begin{lstlisting}[caption={Testbench Verifying Authentication Bug}, label=lst:tb]
module Authentication Bypass_TB;
    reg clk, rst_n, isHashValid;
    reg [127:0] inputHash, correctHash;
    wire authenticationFlag;
Authentication_Bypass uut (
    .clk(clk), .rst_n(rst_n), .isHashValid(isHashValid),
    .inputHash(inputHash),.correctHash(correctHash),
    .authenticationFlag(authenticationFlag)
    );
    initial begin
        $dumpfile("waveform.vcd");
        $dumpvars(0, generated_testbench);
    end
    always #5 clk = ~clk;
    initial begin
        clk = 0; rst_n = 0; isHashValid = 0;
        inputHash = 128'h0;
        correctHash = 
        128'hA5A5A5A5A5A5A5A5A5A5A5A5A5A5A5A5;#5 rst_n = 1;
        #5 clk = 1; #5 clk = 0; isHashValid = 1; 
        inputHash = correctHash;#5 clk = 1;
        #5 clk = 0; isHashValid = 0;#5 clk = 1; 
        inputHash = 128'h0;
        #5 clk = 0; isHashValid = 1; 
        inputHash = 
        128'h5A5A5A5A5A5A5A5A5A5A5A5A5A5A5A5A;
        #5 clk = 1; #5 clk = 0; isHashValid = 0;#5 clk = 1; 
        inputHash = 
        128'h1A1A1A1A1A1A1A1A1A1A1A1A1A1A1A1A;
        #5 clk = 0; inputHash = 128'h0; #5 
        $finish;
    end

    always @(posedge clk) begin
        $strobe("Time=%0t | 
        isHashValid=%b | inputHash=%h | 
        correctHash=%h | authenticationFlag=%b | 
        currentState=%b",
                $time, isHashValid, inputHash, correctHash, 
                authenticationFlag, uut.currentState);
    end
endmodule
\end{lstlisting}
\label{fig:testbench}
\vspace{-0.05in}
\end{figure}

\begin{figure}
    \centering
\begin{tcolorbox}[colback=green!5!white, colframe=green!50!black, title=Validator Output at Timestamp 45]
\scriptsize
\textbf{Expected Region of Interest (ROI):}
\begin{itemize}[leftmargin=*]
  \item \texttt{isHashValid = 1}
  \item \texttt{inputHash =
  128'h5A5A5A5A5A.......}
  \item \texttt{correctHash = 
  128'hA5A5A5A5A5.......}
  \item \texttt{authenticationFlag = 0}
  \item \texttt{currentState = WAIT\_STATE}
\end{itemize}
\textbf{Simulated Monitor:} All values \textbf{matched} the ROI at time 45 ns. FSM reached \texttt{WAIT\_STATE} despite authentication failure.
\vspace{2pt}\textbf{Conclusion:} \textcolor{green!60!black}{\textbf{MATCH}} — Bug validated by signal trace and FSM transition behavior.
\end{tcolorbox}
\caption{Validator output at 45 ns, comparing the expected ROI against simulation results to confirm the authentication‐bypass behavior.}
\label{fig:monitor}
\vspace{-0.2in}
\end{figure}

The generated testbench was validated using the ModelSim simulation environment. The simulation logs captured detailed time-stamped signal transitions, enabling the creation of a structured monitor. Subsequently, the \textit{Bug Validator Agent} analyzed these results, comparing actual signals against the golden monitor predictions. At 45 ns, the agent confirmed the match between simulated and expected outputs, conclusively verifying the FSM’s incorrect transition into WAIT\_STATE under failed authentication, as shown in Figure \ref{fig:monitor}.

\section{Implementation of \textit{SV-LLM}}

\begin{figure}[h!]
\centering
\includegraphics[width=0.5\textwidth]{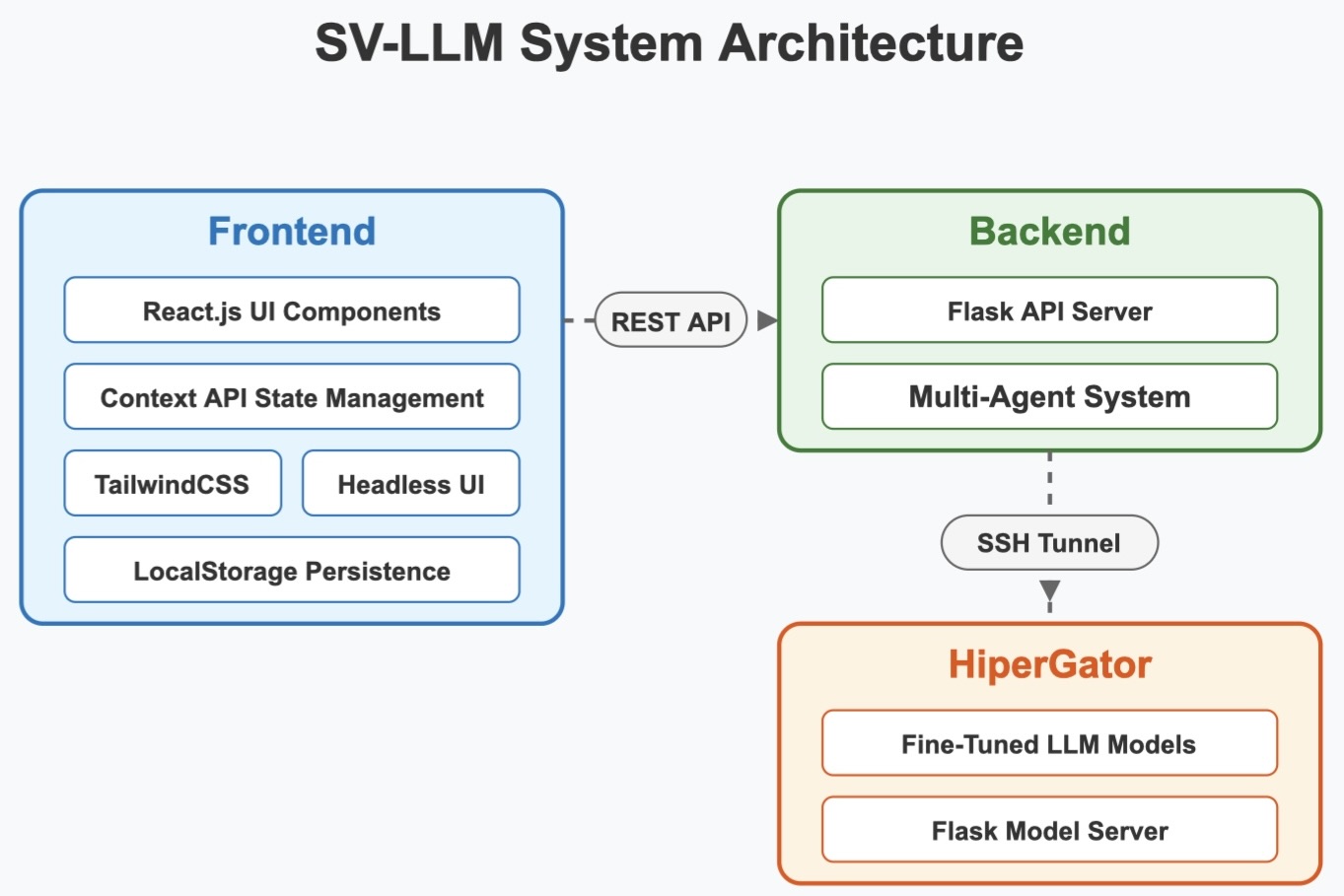}
\caption{\textit{SV-LLM} system architecture}
\label{fig:system-architecture}
\end{figure}

\subsection{Frontend}

The frontend of \textit{SV-LLM}, as illustrated in Figure \ref{fig:system-architecture}, employs a modern React.js technology stack to create a responsive and intuitive user interface for hardware security verification tasks. Built with React.js for component structure, the frontend leverages Context API for comprehensive state management, handling chat history, configuration settings, and theme preferences. The user interface features a chat-based interaction model with specialized markdown rendering for code and technical content, dedicated components to display SVAs, and seamless file upload functionality for hardware design specifications. The UI is styled using TailwindCSS with a customized design system that includes both light and dark modes for extended working sessions, while Headless UI components ensure accessibility and consistent interaction patterns across the application.

The frontend architecture implements several security-specific features critical for hardware verification workflows. It includes contextual input handling with specialized forms that appear when additional information is needed (such as design files or vulnerability specifications), a sophisticated SVA display component that offers syntax highlighting and direct download of generated assertions, and persistent conversation history maintained in browser local storage for privacy. The system supports multiple configurable LLM models through a settings panel, real-time feedback mechanisms for response quality, and granular context window adjustments to optimize token usage. This interface design enables hardware security experts to interact with complex verification capabilities through an accessible, purpose-built environment that abstracts the underlying complexity while maintaining all relevant technical context.

\subsection{Backend}
The \textit{SV-LLM} system employs a distributed architecture in which the React.js front-end interacts with the Flask back-end through a RESTful API that handles user queries and returns structured responses. The frontend makes asynchronous HTTP requests to the backend's endpoints, transmitting user messages, file uploads, and configuration settings while maintaining stateless communication patterns in accordance with REST principles. On the server side, the Flask backend connects to specialized fine-tuned models hosted on the University of Florida's HiperGator supercomputing infrastructure through a secure SSH tunnel. This tunnel creates an encrypted channel between the backend server and HiperGator compute nodes, enabling reliable, high-performance model inference without requiring direct Internet exposure of the supercomputer's resources. The SSH tunneling mechanism allows the system to leverage HiperGator's computational capabilities for resource-intensive model operations while maintaining security and integration with the lightweight Flask API serving the frontend.

\section{Result Analysis} 

\subsection{Detection of Security Vulnerability}

\begin{figure}[t]
\centering
\includegraphics[scale=.36]{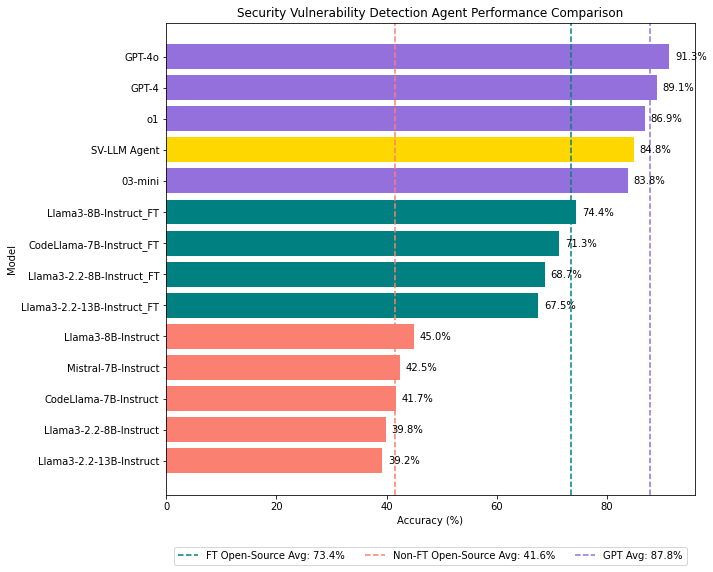}
\caption{Comparison of \textit{Security Vulnerability Detection Agent} with other Proprietary and Open-source LLMs}
\label{fig:result_bug_detection}
\end{figure}

The detection accuracy results for proprietary, fine-tuned, and non-fine-tuned open-source models are presented in Figure~\ref{fig:result_bug_detection}. Among these, the \textit{Security Vulnerability Detection Agent}, a fine-tuned Mistral-7B-Instruct model developed as part of the \textit{SV-LLM} framework, achieved a detection accuracy of 84.8\%, significantly outperforming its non-fine-tuned counterpart, which achieved only 42.5\%. This 42.3 percentage point improvement underscores the effectiveness of our domain-specific fine-tuning strategy in equipping open-source LLMs with specialized RTL security reasoning capabilities.

While proprietary models such as \textit{GPT-4o} (91.3\%) and \textit{o1} (86.9\%) lead in absolute accuracy, their closed-source nature and resource demands make them less practical for widespread deployment. In contrast, the \textit{Security Vulnerability Detection Agent} offers a transparent, cost-efficient alternative that approaches proprietary-level performance.

Other fine-tuned open-source models, such as \textit{Llama-3.1-8B} and \textit{Llama-3.2-3B}, also show improved accuracy after fine-tuning, reaching 74.4\% and 68.7\% respectively. This trend highlights the value of domain adaptation. Conversely, non-fine-tuned models average around 40\% accuracy, confirming that general-purpose LLMs lack the RTL-specific knowledge required for effective security vulnerability detection.

Model capacity also plays a role—larger models generally yield higher accuracy post-fine-tuning. However, the \textit{Security Vulnerability Detection Agent} demonstrates that parameter-efficient fine-tuning on a relatively small open-source model can still yield strong performance while maintaining computational efficiency.

These findings validate the \textit{SV-LLM} framework’s design goal: enabling accurate, scalable, and explainable RTL security verification using tailored LLM agents. The success of the \textit{Security Vulnerability Detection Agent} reinforces the potential of fine-tuned open-source models to serve as practical and robust components in automated hardware security analysis pipelines.

\subsection{Bug Vallidation}

 To evaluate the efficacy of the proposed methodology for the bug validation agent, extensive experiments were conducted on a diverse set of RTL designs incorporating various types of security vulnerabilities. The primary metric employed was the bug-validated testbench generation rate, representing the proportion of testbenches that successfully triggered and validated the intended vulnerabilities.

The experimental results shown in Figure \ref{fig:bug_detection} clearly demonstrate the superior performance of our proposed framework compared to the zero-shot prompting approaches. Specifically, our agent-driven framework achieved significantly higher bug validation rates: 87\% with the GPT-4o model, 82\% with the o1 model, and 89\% with the o3-mini model. In stark contrast, the baseline zero-shot prompting methods yielded considerably lower bug validation rates, with GPT-4o at 18\%, o1 at 20\%, and o3-mini at 43\%. This performance disparity underscores the effectiveness of the structured, iterative refinement, and targeted prompting strategies employed by our agent framework.

\begin{figure}[t]
\centering
\includegraphics[scale=.205]{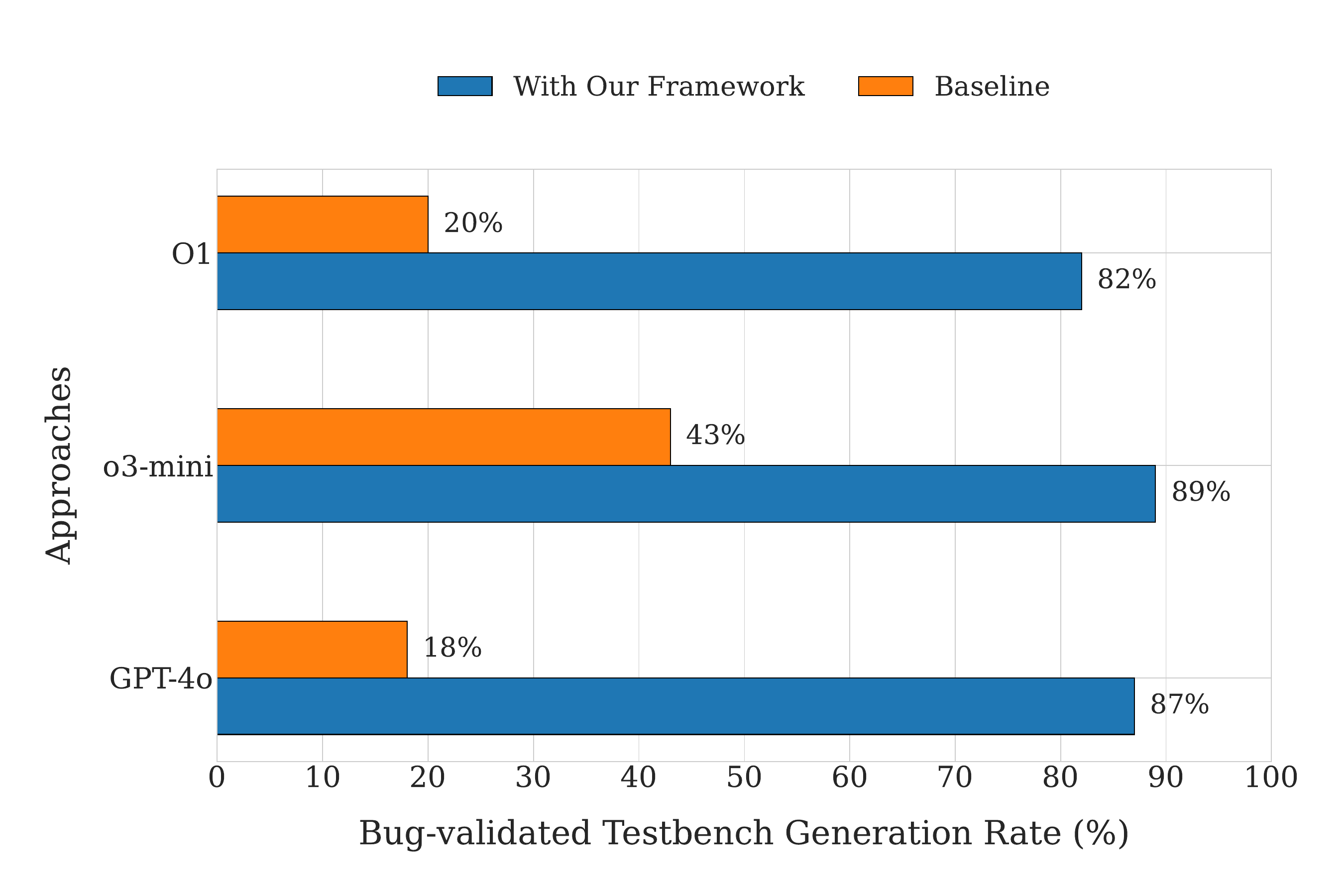}
\caption{Comparison of bug-validated testbench generation rates across different approaches. Our Work consistently outperforms baseline prompting methods for each LLM backend, demonstrating significant improvements in security-focused RTL validation.}
\label{fig:bug_detection}
\end{figure}

The observed high validation rates underscore the robustness and model-agnostic nature of the agent framework, emphasizing its capability to consistently produce accurate, executable testbenches capable of precisely validating complex security vulnerabilities across different LLM configurations. This outcome clearly addresses and overcomes the limitations of simplistic direct-prompting strategies, reinforcing the practical value and reliability of the proposed agentic validation pipeline.

\begin{table*}[!tp]
\vspace{-5pt}
\caption{Security verification capability compared between different methods.}
\vspace{-5pt}
\label{tab: capability}
\centering
%\footnotesize
\begin{tabular}{P{0.22\textwidth} P{0.1\textwidth} P{0.1\textwidth}P{0.1\textwidth}P{0.12\textwidth}P{0.1\textwidth}P{0.12\textwidth}}
\toprule
Method & Asset identification & Threat modeling & RTL bug/vuln. detection & Testbench generation & Property generation \\
 %\cmidrule(lr){3-5} 
 %& layer & SVW-1 & SVW-2 to  & SVW-7 \\
 %& & & SVW-6 & \\
\cmidrule(lr){1-1} \cmidrule(lr){2-2} \cmidrule(lr){3-3}\cmidrule(lr){4-4}\cmidrule(lr){5-5} \cmidrule(lr){6-6}  
  Threat Model based Analysis \cite{halak2021cist, rostami2013hardware, di2007hardware} & \textcolor{red}{\ding{55}} & \textcolor{green}{\ding{51}} & \textcolor{red}{\ding{55}} & \textcolor{red}{\ding{55}}  & \textcolor{red}{\ding{55}}\\
 \cmidrule(lr){1-1} \cmidrule(lr){2-2} \cmidrule(lr){3-3}\cmidrule(lr){4-4}\cmidrule(lr){5-5} \cmidrule(lr){6-6} 
  
  Asset Identification based Analysis \cite{ray2015security, ray2017system, contreras2017security, nahiyan2020script, basak2015flexible, meza2023security} & \textcolor{green}{\ding{51}} & \textcolor{red}{\ding{55}} & \textcolor{red}{\ding{55}} & \textcolor{red}{\ding{55}}  & \textcolor{red}{\ding{55}}\\
 \cmidrule(lr){1-1} \cmidrule(lr){2-2} \cmidrule(lr){3-3}\cmidrule(lr){4-4}\cmidrule(lr){5-5} \cmidrule(lr){6-6} 
 Static RTL verification (e.g. formal\cite{farzana2019soc, hu2016towards, assertion} and/or concolic)\cite{rajendran2016formal,lyu2020scalable} & \textcolor{green}{\ding{51}} & \textcolor{red}{\ding{55}} & \textcolor{green}{\ding{51}} & \textcolor{red}{\ding{55}}  & \textcolor{green}{\ding{51}}\\

 \cmidrule(lr){1-1} \cmidrule(lr){2-2} \cmidrule(lr){3-3}\cmidrule(lr){4-4}\cmidrule(lr){5-5} \cmidrule(lr){6-6}  
 Dynamic RTL verification (e.g. fuzz\cite{tyagi2022thehuzz,laeufer2018rfuzz}, pen testing\cite{al2024re})\cite{lyu2019automated}) & \textcolor{red}{\ding{55}} & \textcolor{red}{\ding{55}} & \textcolor{green}{\ding{51}} & \textcolor{red}{\ding{55}}  & \textcolor{red}{\ding{55}}\\
  \cmidrule(lr){1-1} \cmidrule(lr){2-2} \cmidrule(lr){3-3}\cmidrule(lr){4-4}\cmidrule(lr){5-5} \cmidrule(lr){6-6}   
  ML-based testbench generation\cite{fine2006harnessing,samarah2006automated} & \textcolor{red}{\ding{55}} & \textcolor{red}{\ding{55}} & \textcolor{red}{\ding{55}} & \textcolor{green}{\ding{51}}  & \textcolor{red}{\ding{55}}\\
  
  \cmidrule(lr){1-1} \cmidrule(lr){2-2} \cmidrule(lr){3-3}\cmidrule(lr){4-4}\cmidrule(lr){5-5} \cmidrule(lr){6-6} 
     \textit{SV-LLM} (this work) & \textcolor{green}{\ding{51}} & \textcolor{green}{\ding{51}} & \textcolor{green}{\ding{51}} & \textcolor{green}{\ding{51}}  & \textcolor{green}{\ding{51}}\\

\bottomrule
\end{tabular}
\end{table*}
\section{Related Works \& Comparisons}\label{sec: related_works}

Related works can be broadly categorized into two dimensions: (i) traditional, non-LLM-based approaches applied to various aspects of hardware security verification, and (ii) emerging LLM-based techniques recently introduced in this domain. In the following discussion, we analyze each category in turn, aligning them with the six verification dimensions addressed by \textit{SV-LLM}.

\subsection{Traditional verification approaches}\label{subsec: trad_verif_approach}
Asset identification in hardware security has been explored through various methodologies to leverage security‑critical elements for policy enforcement and vulnerability analysis. In \cite{ray2015security} and \cite{basak2015flexible}, the authors leverage designated hardware security assets to enforce security policies directly within the design flow. In \cite{contreras2017security} and \cite{nahiyan2020script}, designers analyze confidentiality and integrity violations by inserting DfT logic and evaluating power side‑channel leakage based on selected security assets. Ray et al. \cite{ray2017system} propose an automated framework for identifying security assets across diverse threat scenarios and exploring corresponding adversarial exploits. The study in \cite{meza2023security} performs a comprehensive security verification of the OpenTitan hardware root of trust, including systematic identification of its critical security assets.

Recent efforts in hardware security have focused on developing structured threat modeling frameworks to assess and mitigate vulnerabilities in different stages of the hardware lifecycle. Halak et al. \cite{halak2021cist} introduced Cist, a lifecycle-wide threat modeling framework for hardware supply chain security that systematizes emerging attacks and defenses and validates countermeasures through application‑specific case studies. Rostami et al. \cite{rostami2013hardware} proposed comprehensive hardware threat models and quantitative security metrics to assess circuit resilience against malicious modifications and to enable systematic comparisons of defense techniques. Di and Smith\cite{di2007hardware} developed a threat modeling methodology for integrated circuits that characterizes potential malicious logic insertions and guides checking tools to assess the trustworthiness of the IC.

There have been several different approaches for vulnerability detection at the RTL stage.
Machine learning-based hardware verification techniques \cite{elnaggar2018machine} often encounter limitations stemming from design dependency and data scarcity, which restrict their generalizability and effectiveness. Recently, dynamic verification techniques such as fuzzing \cite{chen2023hypfuzz, laeufer2018rfuzz, tyagi2022thehuzz} and penetration testing \cite{al2023sharpen, al2024re} have gained attention for security verification at the RTL level. In parallel, Concolic testing \cite{lyu2019automated, lyu2020scalable}, a static analysis approach, has also been applied for security validation in this context. However, in contrast to the \textit{Security Vulnerability Detection Agent} integrated within \textit{SV-LLM}, these techniques typically require varying levels of expert manual intervention, introducing a higher likelihood of human error and increasing the overall verification effort and time.

Several earlier works proposed automated testbench generation using genetic algorithms \cite{lajolo2000automatic, shen2008coverage, samarah2006automated} and feedback-driven, coverage-directed techniques that integrate machine learning to bias stimulus generation toward coverage gaps. 
These approaches operate under the premise that learning mechanisms can effectively analyze existing test data and coverage metrics to guide the generation of new stimuli aimed at achieving comprehensive coverage. 
Among these, \cite{fine2006harnessing, baras2011automatic} employ Bayesian networks to improve the coverage efficacy of automatically generated testbenches. However, unlike the Security Bug Validation agent in \textit{SV-LLM}, such approaches focused solely on functional coverage and did not address security bug validation.

Recent property‑driven hardware security approaches have focused on specifying and verifying security properties within standard hardware design workflows. Hu et al. \cite{hu2016towards} introduced a novel property specification language to enforce information flow and statistical security properties, enabling their translation and verification using existing hardware design tools. Farzana et al. \cite{farzana2019soc} developed a comprehensive set of reusable, architecture‐agnostic security properties and derived quantitative metrics to guide security‐aware design rule enforcement in SoC verification. Witharana et al. \cite{assertion} proposed an automated framework for generating tailored security assertions that streamline vulnerability‐specific verification in complex SoC designs.

A summary of these comparisons is presented in Table~\ref{tab: capability}. As illustrated in the table, \textit{SV-LLM} demonstrates a uniquely comprehensive scope, being the only approach that supports all five critical verification dimensions. 
Beyond this functional breadth, \textit{SV-LLM} further offers significant advantages in automation and real-time interactivity (provided through its chat agent), substantially reducing reliance on manual expertise and accelerating the overall verification process.

\subsection{LLM-based methods in hardware security}\label{subsec: llm_based_methods}

LLMs have seen increasing adoption in hardware verification, particularly for tasks related to security analysis and validation. For example, Saha et al. \cite{saha2024llm} investigated the applicability of LLMs across a range of SoC security tasks and highlighted key limitations, such as the difficulty of processing large hardware designs due to restricted context length. In a related effort, Bhunia et al. \cite{paria2023divas} used LLM to identify security vulnerabilities in SoC designs, map them to relevant CWEs, generate corresponding assertions, and enforce security policies, demonstrating efficacy on open-source SoC benchmarks. Fu et al. \cite{10409307} curated a dataset of RTL defects and their remediation from open-source projects, training medium-sized LLMs to detect bugs; however, large RTL files were omitted due to token limitations. Additionally, Saha et al. \cite{secrtllm} used prompt engineering to uncover 16 distinct security vulnerabilities in FSM designs.

In parallel, several specialized LLM-based approaches have emerged, each targeting specific aspects of hardware verification. RTL code debugging frameworks such as UVLLM \cite{uvllm}, RTLFixer \cite{rtlfixer}, HDLDebugger \cite{hdldebugger}, LLM4DV \cite{llm4dv}, VeriAssist \cite{veriassist}, and VeriDebug \cite{veridebug} leverage LLMs to detect and explain RTL-level issues, including syntactic, structural, and semantic inconsistencies. Other frameworks focus on identifying vulnerabilities in RTL designs, including SoCureLLM \cite{socurellm}, SecRTL-LLM \cite{secrtllm}, and BugWhisperer \cite{bugwhisperer}. Assertion-based verification techniques have also used LLMs to automatically generate functional and security properties, typically encoded as SVAs, that enable formal verification against intended behaviors and security requirements \cite{flag, assertionforge, fveval, nl2sva, assertionbench, 10546729, kande2024security, ayalasomayajula2024lasp}. In addition, frameworks such as ThreatLens \cite{saha2025threatlens} have focused on LLM-driven threat modeling and policy generation to inform secure hardware design and validation practices.

Again, we want to note that none of these approaches offer the same versatility and breadth of applicability as \textit{SV-LLM}.

\section{Conclusion}
\label{sec:conclusion}

In conclusion, this work has introduced \textit{SV‑LLM}, a novel multi‑agent AI assistant framework that leverages large language models to automate key stages of SoC security verification. By decomposing the workflow into specialized agents, responsible for answering security verification questions, security asset identification, threat modeling, property generation, vulnerability detection, and simulation‑based bug validation, \textit{SV‑LLM} effectively reduces manual effort and accelerates the discovery and mitigation of security flaws early in the design cycle. Our case studies on representative SoC designs demonstrate that \textit{SV‑LLM} not only streamlines the verification process but also enhances coverage and accuracy compared to traditional, manual methods. Beyond its immediate improvements in efficiency and scalability, \textit{SV‑LLM} lays the groundwork for a more integrated approach to hardware security. We believe that \textit{SV‑LLM} represents a significant step toward fully automated, intelligence‑driven security verification, capable of evolving alongside the increasing complexity of modern SoC architectures.

\bibliographystyle{IEEEtran}
\bibliography{IEEEabrv,refs}

\begin{thebibliography}{10}
\providecommand{\url}[1]{#1}
\csname url@samestyle\endcsname
\providecommand{\newblock}{\relax}
\providecommand{\bibinfo}[2]{#2}
\providecommand{\BIBentrySTDinterwordspacing}{\spaceskip=0pt\relax}
\providecommand{\BIBentryALTinterwordstretchfactor}{4}
\providecommand{\BIBentryALTinterwordspacing}{\spaceskip=\fontdimen2\font plus
\BIBentryALTinterwordstretchfactor\fontdimen3\font minus \fontdimen4\font\relax}
\providecommand{\BIBforeignlanguage}[2]{{%
\expandafter\ifx\csname l@#1\endcsname\relax
\typeout{** WARNING: IEEEtran.bst: No hyphenation pattern has been}%
\typeout{** loaded for the language `#1'. Using the pattern for}%
\typeout{** the default language instead.}%
\else
\language=\csname l@#1\endcsname
\fi
#2}}
\providecommand{\BIBdecl}{\relax}
\BIBdecl

\bibitem{darbari2024verification}
\BIBentryALTinterwordspacing
A.~Darbari. (2024, April) Verification in crisis. Accessed: 2025-04-25. [Online]. Available: \url{https://semiengineering.com/verification-in-crisis/}
\BIBentrySTDinterwordspacing

\bibitem{ravichandran2022pacman}
J.~Ravichandran, W.~T. Na, J.~Lang, and M.~Yan, ``Pacman: attacking arm pointer authentication with speculative execution,'' in \emph{Proceedings of the 49th Annual International Symposium on Computer Architecture}, 2022, pp. 685--698.

\bibitem{vicarte2022augury}
J.~S. Vicarte, M.~Flanders, R.~Paccagnella, G.~Garrett-Grossman, A.~Morrison, C.~Fletcher, and D.~Kohlbrenner, ``Augury: Using data memory-dependent prefetchers to leak data at rest,'' in \emph{2022 IEEE Symposium on Security and Privacy (SP)}.\hskip 1em plus 0.5em minus 0.4em\relax IEEE Computer Society, 2022, pp. 1518--1518.

\bibitem{thomas2024riscvuzz}
F.~Thomas, L.~Hetterich, R.~Zhang, D.~Weber, L.~Gerlach, and M.~Schwarz, ``Riscvuzz: Discovering architectural cpu vulnerabilities via differential hardware fuzzing,'' 2024.

\bibitem{rajendran2016formal}
J.~Rajendran, A.~M. Dhandayuthapany, V.~Vedula, and R.~Karri, ``Formal security verification of third party intellectual property cores for information leakage,'' in \emph{2016 29th International conference on VLSI design and 2016 15th international conference on embedded systems (VLSID)}.\hskip 1em plus 0.5em minus 0.4em\relax IEEE, 2016, pp. 547--552.

\bibitem{subramanyan2014formal}
P.~Subramanyan and D.~Arora, ``Formal verification of taint-propagation security properties in a commercial soc design,'' in \emph{2014 Design, Automation \& Test in Europe Conference \& Exhibition (DATE)}.\hskip 1em plus 0.5em minus 0.4em\relax IEEE, 2014, pp. 1--2.

\bibitem{nahiyan2017hardware}
A.~Nahiyan, M.~Sadi, R.~Vittal, G.~Contreras, D.~Forte, and M.~Tehranipoor, ``Hardware trojan detection through information flow security verification,'' in \emph{2017 IEEE International Test Conference (ITC)}.\hskip 1em plus 0.5em minus 0.4em\relax IEEE, 2017, pp. 1--10.

\bibitem{lyu2020scalable}
Y.~Lyu and P.~Mishra, ``Scalable concolic testing of rtl models,'' \emph{IEEE Transactions on Computers}, vol.~70, no.~7, pp. 979--991, 2020.

\bibitem{lyu2019automated}
Y.~Lyu, A.~Ahmed, and P.~Mishra, ``Automated activation of multiple targets in rtl models using concolic testing,'' in \emph{2019 Design, Automation \& Test in Europe Conference \& Exhibition (DATE)}.\hskip 1em plus 0.5em minus 0.4em\relax IEEE, 2019, pp. 354--359.

\bibitem{kibria2022rtl}
R.~Kibria, M.~S. Rahman, F.~Farahmandi, and M.~Tehranipoor, ``Rtl-fsmx: Fast and accurate finite state machine extraction at the rtl for security applications,'' in \emph{2022 IEEE International Test Conference (ITC)}.\hskip 1em plus 0.5em minus 0.4em\relax IEEE, 2022, pp. 165--174.

\bibitem{al2023quardtropy}
H.~Al~Shaikh, M.~B. Monjil, K.~Z. Azar, F.~Farahmandi, M.~Tehranipoor, and F.~Rahman, ``Quardtropy: Detecting and quantifying unauthorized information leakage in hardware designs using g-entropy,'' in \emph{2023 IEEE International Symposium on Defect and Fault Tolerance in VLSI and Nanotechnology Systems (DFT)}.\hskip 1em plus 0.5em minus 0.4em\relax IEEE, 2023, pp. 1--6.

\bibitem{gohil2024mabfuzz}
V.~Gohil, R.~Kande, C.~Chen, A.-R. Sadeghi, and J.~Rajendran, ``Mabfuzz: Multi-armed bandit algorithms for fuzzing processors,'' in \emph{2024 Design, Automation \& Test in Europe Conference \& Exhibition (DATE)}.\hskip 1em plus 0.5em minus 0.4em\relax IEEE, 2024, pp. 1--6.

\bibitem{hossain2023socfuzzer}
M.~M. Hossain, A.~Vafaei, K.~Z. Azar, F.~Rahman, F.~Farahmandi, and M.~Tehranipoor, ``Socfuzzer: Soc vulnerability detection using cost function enabled fuzz testing,'' in \emph{2023 Design, Automation \& Test in Europe Conference \& Exhibition (DATE)}.\hskip 1em plus 0.5em minus 0.4em\relax IEEE, 2023, pp. 1--6.

\bibitem{azar2022fuzz}
K.~Z. Azar, M.~M. Hossain, A.~Vafaei, H.~Al~Shaikh, N.~N. Mondol, F.~Rahman, M.~Tehranipoor, and F.~Farahmandi, ``Fuzz, penetration, and ai testing for soc security verification: Challenges and solutions,'' \emph{Cryptology ePrint Archive}, 2022.

\bibitem{trippel2021fuzzing}
T.~Trippel, K.~G. Shin, A.~Chernyakhovsky, G.~Kelly, D.~Rizzo, and M.~Hicks, ``Fuzzing hardware like software,'' \emph{arXiv preprint arXiv:2102.02308}, 2021.

\bibitem{al2023sharpen}
H.~Al-Shaikh, A.~Vafaei, M.~M.~M. Rahman, K.~Z. Azar, F.~Rahman, F.~Farahmandi, and M.~Tehranipoor, ``Sharpen: Soc security verification by hardware penetration test,'' in \emph{Proceedings of the 28th Asia and South Pacific Design Automation Conference}, 2023, pp. 579--584.

\bibitem{al2024re}
H.~Al~Shaikh, S.~Saha, K.~Z. Azar, F.~Farahmandi, M.~Tehranipoor, and F.~Rahman, ``Re-pen: Reinforcement learning-enforced penetration testing for soc security verification,'' \emph{IEEE Transactions on Very Large Scale Integration (VLSI) Systems}, no.~01, pp. 1--14, 2024.

\bibitem{tarek2023benchmarking}
S.~Tarek, H.~Al~Shaikh, S.~R. Rajendran, and F.~Farahmandi, ``Benchmarking of soc-level hardware vulnerabilities: A complete walkthrough,'' in \emph{2023 IEEE Computer Society Annual Symposium on VLSI (ISVLSI)}.\hskip 1em plus 0.5em minus 0.4em\relax IEEE, 2023, pp. 1--6.

\bibitem{saha2024llm}
D.~Saha, S.~Tarek, K.~Yahyaei, S.~K. Saha, J.~Zhou, M.~Tehranipoor, and F.~Farahmandi, ``Llm for soc security: A paradigm shift,'' \emph{IEEE Access}, vol.~12, pp. 155\,498--155\,521, 2024.

\bibitem{kande2024security}
R.~Kande, H.~Pearce, B.~Tan, B.~Dolan-Gavitt, S.~Thakur, R.~Karri, and J.~Rajendran, ``(security) assertions by large language models,'' \emph{IEEE Transactions on Information Forensics and Security}, 2024.

\bibitem{ahmad2024hardware}
B.~Ahmad, S.~Thakur, B.~Tan, R.~Karri, and H.~Pearce, ``On hardware security bug code fixes by prompting large language models,'' \emph{IEEE Transactions on Information Forensics and Security}, 2024.

\bibitem{meng2024nspg}
X.~Meng, A.~Srivastava, A.~Arunachalam, A.~Ray, P.~H. Silva, R.~Psiakis, Y.~Makris, and K.~Basu, ``Nspg: Natural language processing-based security property generator for hardware security assurance,'' in \emph{Proceedings of the 61st ACM/IEEE Design Automation Conference}, 2024, pp. 1--6.

\bibitem{secrtllm}
D.~Saha, K.~Yahyaei, S.~K. Saha, M.~Tehranipoor, and F.~Farahmandi, ``Empowering hardware security with llm: The development of a vulnerable hardware database,'' in \emph{2024 IEEE International Symposium on Hardware Oriented Security and Trust (HOST)}.\hskip 1em plus 0.5em minus 0.4em\relax IEEE, 2024, pp. 233--243.

\bibitem{socurellm}
S.~Tarek, D.~Saha, S.~K. Saha, M.~Tehranipoor, and F.~Farahmandi, ``Socurellm: An llm-driven approach for large-scale system-on-chip security verification and policy generation,'' \emph{Cryptology ePrint Archive}, 2024.

\bibitem{saha2025threatlens}
D.~Saha, H.~Al~Shaikh, S.~Tarek, and F.~Farahmandi, ``Special session: Threatlens: Llm-guided threat modeling and test plan generation for hardware security verification,'' in \emph{2025 IEEE 43rd VLSI Test Symposium (VTS)}, 2025, pp. 1--5.

\bibitem{bugwhisperer}
S.~Tarek, D.~Saha, S.~K. Saha, and F.~Farahmandi, ``Bugwhisperer: Fine-tuning llms for soc hardware vulnerability detection,'' in \emph{2025 IEEE 43rd VLSI Test Symposium (VTS)}, 2025, pp. 1--5.

\bibitem{paria2024navigating}
S.~Paria, A.~Dasgupta, and S.~Bhunia, ``Navigating soc security landscape on llm-guided paths,'' in \emph{Proceedings of the Great Lakes Symposium on VLSI 2024}, 2024, pp. 252--257.

\bibitem{akyash2024self}
M.~Akyash and H.~M. Kamali, ``Self-hwdebug: Automation of llm self-instructing for hardware security verification,'' in \emph{2024 IEEE Computer Society Annual Symposium on VLSI (ISVLSI)}.\hskip 1em plus 0.5em minus 0.4em\relax IEEE, 2024, pp. 391--396.

\bibitem{10691745}
R.~Afsharmazayejani, M.~M. Shahmiri, P.~Link, H.~Pearce, and B.~Tan, ``Toward hardware security benchmarking of llms,'' in \emph{2024 IEEE LLM Aided Design Workshop (LAD)}, 2024, pp. 1--7.

\bibitem{faruque2024trojanwhisper}
M.~O. Faruque, P.~Jamieson, A.~Patooghy, and A.-H.~A. Badawy, ``Trojanwhisper: Evaluating pre-trained llms to detect and localize hardware trojans,'' \emph{arXiv preprint arXiv:2412.07636}, 2024.

\bibitem{latibari2024automated}
B.~S. Latibari, S.~Ghimire, M.~A. Chowdhury, N.~Nazari, K.~I. Gubbi, H.~Homayoun, A.~Sasan, and S.~Salehi, ``Automated hardware logic obfuscation framework using gpt,'' in \emph{2024 IEEE 17th Dallas Circuits and Systems Conference (DCAS)}.\hskip 1em plus 0.5em minus 0.4em\relax IEEE, 2024, pp. 1--5.

\bibitem{10904479}
V.~T. Hayashi and W.~Vicente~Ruggiero, ``Hardware trojan detection in open-source hardware designs using machine learning,'' \emph{IEEE Access}, vol.~13, pp. 37\,771--37\,788, 2025.

\bibitem{11022798}
A.~Menon, S.~S. Miftah, A.~Srivastava, S.~Kundu, S.~Kundu, A.~Raha, S.~Banerjee, D.~Mathaikutty, and K.~Basu, ``Openassert: Towards secure assertion generation using large language models,'' in \emph{2025 IEEE 43rd VLSI Test Symposium (VTS)}, 2025, pp. 1--5.

\bibitem{qiu2024llm}
J.~Qiu, K.~Lam, G.~Li, A.~Acharya, T.~Y. Wong, A.~Darzi, W.~Yuan, and E.~J. Topol, ``Llm-based agentic systems in medicine and healthcare,'' \emph{Nature Machine Intelligence}, vol.~6, no.~12, pp. 1418--1420, 2024.

\bibitem{acharya2025agentic}
D.~B. Acharya, K.~Kuppan, and B.~Divya, ``Agentic ai: Autonomous intelligence for complex goals--a comprehensive survey,'' \emph{IEEE Access}, 2025.

\bibitem{lei2024macm}
B.~Lei, Y.~Zhang, S.~Zuo, A.~Payani, and C.~Ding, ``Macm: Utilizing a multi-agent system for condition mining in solving complex mathematical problems,'' \emph{arXiv preprint arXiv:2404.04735}, 2024.

\bibitem{zeng2024autodefense}
Y.~Zeng, Y.~Wu, X.~Zhang, H.~Wang, and Q.~Wu, ``Autodefense: Multi-agent llm defense against jailbreak attacks,'' \emph{arXiv preprint arXiv:2403.04783}, 2024.

\bibitem{yu2024fincon}
Y.~Yu, Z.~Yao, H.~Li, Z.~Deng, Y.~Jiang, Y.~Cao, Z.~Chen, J.~Suchow, Z.~Cui, R.~Liu \emph{et~al.}, ``Fincon: A synthesized llm multi-agent system with conceptual verbal reinforcement for enhanced financial decision making,'' \emph{Advances in Neural Information Processing Systems}, vol.~37, pp. 137\,010--137\,045, 2024.

\bibitem{manish2024autonomous}
S.~Manish, ``An autonomous multi-agent llm framework for agile software development,'' \emph{International Journal of Trend in Scientific Research and Development}, vol.~8, no.~5, pp. 892--898, 2024.

\bibitem{chen2024survey}
S.~Chen, Y.~Liu, W.~Han, W.~Zhang, and T.~Liu, ``A survey on multi-generative agent system: Recent advances and new frontiers,'' \emph{arXiv preprint arXiv:2412.17481}, 2024.

\bibitem{sa_edi}
\BIBentryALTinterwordspacing
(2021) Security annotation for electronic design integration standard. [Online]. Available: \url{https://www.accellera.org/images/downloads/standards/Accellera\_SA-EDI\_Standard\_v10.pdf}
\BIBentrySTDinterwordspacing

\bibitem{ieee_p3164}
``Asset identification for electronic design ip,'' \emph{Asset Identification for Electronic Design IP}, pp. 1--26, 2024.

\bibitem{nath2025toward}
S.~K.~D. Nath and B.~Tan, ``Toward automated potential primary asset identification in verilog designs,'' \emph{arXiv preprint arXiv:2502.04648}, 2025.

\bibitem{ayalasomayajula2024automatic}
A.~Ayalasomayajula, N.~F. Dipu, M.~M. Tehranipoor, and F.~Farahmandi, ``Automatic asset identification for assertion-based soc security verification,'' \emph{IEEE Transactions on Computer-Aided Design of Integrated Circuits and Systems}, vol.~43, no.~10, pp. 3264--3277, 2024.

\bibitem{ayalasomayajula2024lasp}
A.~Ayalasomayajula, R.~Guo, J.~Zhou, S.~K. Saha, and F.~Farahmandi, ``Lasp: Llm assisted security property generation for soc verification,'' in \emph{Proceedings of the 2024 ACM/IEEE International Symposium on Machine Learning for CAD}, 2024, pp. 1--7.

\bibitem{laeufer2018rfuzz}
K.~Laeufer, J.~Koenig, D.~Kim, J.~Bachrach, and K.~Sen, ``Rfuzz: Coverage-directed fuzz testing of rtl on fpgas,'' in \emph{2018 IEEE/ACM International Conference on Computer-Aided Design (ICCAD)}.\hskip 1em plus 0.5em minus 0.4em\relax IEEE, 2018, pp. 1--8.

\bibitem{tyagi2022thehuzz}
A.~Tyagi, A.~Crump, A.-R. Sadeghi, G.~Persyn, J.~Rajendran, P.~Jauernig, and R.~Kande, ``Thehuzz: Instruction fuzzing of processors using golden-reference models for finding software-exploitable vulnerabilities,'' \emph{arXiv preprint arXiv:2201.09941}, 2022.

\bibitem{dipu2024formalfuzzer}
N.~F. Dipu, M.~M. Hossain, K.~Z. Azar, F.~Farahmandi, and M.~Tehranipoor, ``Formalfuzzer: Formal verification assisted fuzz testing for soc vulnerability detection,'' in \emph{2024 29th Asia and South Pacific Design Automation Conference (ASP-DAC)}.\hskip 1em plus 0.5em minus 0.4em\relax IEEE, 2024, pp. 355--361.

\bibitem{halak2021cist}
B.~Halak, ``Cist: A threat modelling approach for hardware supply chain security,'' \emph{Hardware Supply Chain Security: Threat Modelling, Emerging Attacks and Countermeasures}, pp. 3--65, 2021.

\bibitem{rostami2013hardware}
M.~Rostami, F.~Koushanfar, J.~Rajendran, and R.~Karri, ``Hardware security: Threat models and metrics,'' in \emph{2013 IEEE/ACM International Conference on Computer-Aided Design (ICCAD)}.\hskip 1em plus 0.5em minus 0.4em\relax IEEE, 2013, pp. 819--823.

\bibitem{di2007hardware}
J.~Di and S.~Smith, ``A hardware threat modeling concept for trustable integrated circuits,'' in \emph{2007 IEEE Region 5 Technical Conference}.\hskip 1em plus 0.5em minus 0.4em\relax IEEE, 2007, pp. 354--357.

\bibitem{ray2015security}
S.~Ray and Y.~Jin, ``Security policy enforcement in modern soc designs,'' in \emph{2015 IEEE/ACM International Conference on Computer-Aided Design (ICCAD)}.\hskip 1em plus 0.5em minus 0.4em\relax IEEE, 2015, pp. 345--350.

\bibitem{ray2017system}
S.~Ray, E.~Peeters, M.~M. Tehranipoor, and S.~Bhunia, ``System-on-chip platform security assurance: Architecture and validation,'' \emph{Proceedings of the IEEE}, vol. 106, no.~1, pp. 21--37, 2017.

\bibitem{contreras2017security}
G.~K. Contreras, A.~Nahiyan, S.~Bhunia, D.~Forte, and M.~Tehranipoor, ``Security vulnerability analysis of design-for-test exploits for asset protection in socs,'' in \emph{2017 22nd Asia and South Pacific Design Automation Conference (ASP-DAC)}.\hskip 1em plus 0.5em minus 0.4em\relax IEEE, 2017, pp. 617--622.

\bibitem{nahiyan2020script}
A.~Nahiyan, J.~Park, M.~He, Y.~Iskander, F.~Farahmandi, D.~Forte, and M.~Tehranipoor, ``Script: A cad framework for power side-channel vulnerability assessment using information flow tracking and pattern generation,'' \emph{ACM Transactions on Design Automation of Electronic Systems (TODAES)}, vol.~25, no.~3, pp. 1--27, 2020.

\bibitem{basak2015flexible}
A.~Basak, S.~Bhunia, and S.~Ray, ``A flexible architecture for systematic implementation of soc security policies,'' in \emph{2015 IEEE/ACM International Conference on Computer-Aided Design (ICCAD)}.\hskip 1em plus 0.5em minus 0.4em\relax IEEE, 2015, pp. 536--543.

\bibitem{meza2023security}
A.~Meza, F.~Restuccia, J.~Oberg, D.~Rizzo, and R.~Kastner, ``Security verification of the opentitan hardware root of trust,'' \emph{IEEE Security \& Privacy}, vol.~21, no.~3, pp. 27--36, 2023.

\bibitem{farzana2019soc}
N.~Farzana, F.~Rahman, M.~Tehranipoor, and F.~Farahmandi, ``Soc security verification using property checking,'' in \emph{2019 IEEE International Test Conference (ITC)}.\hskip 1em plus 0.5em minus 0.4em\relax IEEE, 2019, pp. 1--10.

\bibitem{hu2016towards}
W.~Hu, A.~Althoff, A.~Ardeshiricham, and R.~Kastner, ``Towards property driven hardware security,'' in \emph{2016 17th International Workshop on Microprocessor and SOC Test and Verification (MTV)}.\hskip 1em plus 0.5em minus 0.4em\relax IEEE, 2016, pp. 51--56.

\bibitem{assertion}
H.~Witharana, A.~Jayasena, A.~Whigham, and P.~Mishra, ``Automated generation of security assertions for rtl models,'' \emph{J. Emerg. Technol. Comput. Syst.}, vol.~19, 2023.

\bibitem{fine2006harnessing}
S.~Fine, A.~Freund, I.~Jaeger, Y.~Mansour, Y.~Naveh, and A.~Ziv, ``Harnessing machine learning to improve the success rate of stimuli generation,'' \emph{IEEE Transactions on Computers}, vol.~55, no.~11, pp. 1344--1355, 2006.

\bibitem{samarah2006automated}
A.~Samarah, A.~Habibi, S.~Tahar, and N.~Kharma, ``Automated coverage directed test generation using a cell-based genetic algorithm,'' in \emph{2006 IEEE International High Level Design Validation and Test Workshop}.\hskip 1em plus 0.5em minus 0.4em\relax IEEE, 2006, pp. 19--26.

\bibitem{elnaggar2018machine}
R.~Elnaggar and K.~Chakrabarty, ``Machine learning for hardware security: Opportunities and risks,'' \emph{Journal of Electronic Testing}, vol.~34, pp. 183--201, 2018.

\bibitem{chen2023hypfuzz}
C.~Chen, R.~Kande \emph{et~al.}, ``Hypfuzz:formal-assisted processor fuzzing,'' in \emph{32nd USENIX Security Symposium (USENIX Security 23)}, 2023, pp. 1361--1378.

\bibitem{lajolo2000automatic}
M.~Lajolo, L.~Lavagno, M.~Rebaudengo, M.~S. Reorda, and M.~Violante, ``Automatic test bench generation for simulation-based validation,'' in \emph{Proceedings of the eighth international workshop on Hardware/software codesign}, 2000, pp. 136--140.

\bibitem{shen2008coverage}
H.~Shen, W.~Wei, Y.~Chen, B.~Chen, and Q.~Guo, ``Coverage directed test generation: Godson experience,'' in \emph{2008 17th Asian Test Symposium}.\hskip 1em plus 0.5em minus 0.4em\relax IEEE, 2008, pp. 321--326.

\bibitem{baras2011automatic}
D.~Baras, S.~Fine, L.~Fournier, D.~Geiger, and A.~Ziv, ``Automatic boosting of cross-product coverage using bayesian networks,'' \emph{International Journal on Software Tools for Technology Transfer}, vol.~13, pp. 247--261, 2011.

\bibitem{paria2023divas}
S.~Paria, A.~Dasgupta, and S.~Bhunia, ``Divas: An llm-based end-to-end framework for soc security analysis and policy-based protection,'' \emph{arXiv preprint arXiv:2308.06932}, 2023.

\bibitem{10409307}
W.~Fu, K.~Yang, R.~G. Dutta, X.~Guo, and G.~Qu, ``Llm4sechw: Leveraging domain-specific large language model for hardware debugging,'' in \emph{2023 Asian Hardware Oriented Security and Trust Symposium (AsianHOST)}, 2023, pp. 1--6.

\bibitem{uvllm}
Y.~Hu, J.~Ye, K.~Xu, J.~Sun, S.~Zhang, X.~Jiao, D.~Pan, J.~Zhou, N.~Wang, W.~Shan \emph{et~al.}, ``Uvllm: An automated universal rtl verification framework using llms,'' \emph{arXiv preprint arXiv:2411.16238}, 2024.

\bibitem{rtlfixer}
Y.~Tsai, M.~Liu, and H.~Ren, ``Rtlfixer: Automatically fixing rtl syntax errors with large language model,'' in \emph{Proceedings of the 61st ACM/IEEE Design Automation Conference}, 2024, pp. 1--6.

\bibitem{hdldebugger}
X.~Yao, H.~Li, T.~H. Chan, W.~Xiao, M.~Yuan, Y.~Huang, L.~Chen, and B.~Yu, ``Hdldebugger: Streamlining hdl debugging with large language models,'' \emph{arXiv preprint arXiv:2403.11671}, 2024.

\bibitem{llm4dv}
Z.~Zhang, G.~Chadwick, H.~McNally, Y.~Zhao, and R.~Mullins, ``Llm4dv: Using large language models for hardware test stimuli generation,'' \emph{arXiv preprint arXiv:2310.04535}, 2023.

\bibitem{veriassist}
H.~Huang, Z.~Lin, Z.~Wang, X.~Chen, K.~Ding, and J.~Zhao, ``Towards llm-powered verilog rtl assistant: Self-verification and self-correction,'' \emph{arXiv preprint arXiv:2406.00115}, 2024.

\bibitem{veridebug}
N.~Wang, B.~Yao, J.~Zhou, Y.~Hu, X.~Wang, N.~Guan, and Z.~Jiang, ``Veridebug: A unified llm for verilog debugging via contrastive embedding and guided correction,'' \emph{arXiv preprint arXiv:2504.19099}, 2025.

\bibitem{flag}
Y.-A. Shih, A.~Lin, A.~Gupta, and S.~Malik, ``Flag: Formal and llm-assisted sva generation for formal specifications of on-chip communication protocols,'' \emph{arXiv preprint arXiv:2504.17226}, 2025.

\bibitem{assertionforge}
Y.~Bai, G.~B. Hamad, S.~Suhaib, and H.~Ren, ``Assertionforge: Enhancing formal verification assertion generation with structured representation of specifications and rtl,'' \emph{arXiv preprint arXiv:2503.19174}, 2025.

\bibitem{fveval}
M.~Kang, M.~Liu, G.~B. Hamad, S.~Suhaib, and H.~Ren, ``Fveval: Understanding language model capabilities in formal verification of digital hardware,'' \emph{arXiv preprint arXiv:2410.23299}, 2024.

\bibitem{nl2sva}
C.~Sun, C.~Hahn, and C.~Trippel, ``Towards improving verification productivity with circuit-aware translation of natural language to systemverilog assertions,'' in \emph{First International Workshop on Deep Learning-aided Verification}, 2023.

\bibitem{assertionbench}
V.~Pulavarthi, D.~Nandal, S.~Dan, and D.~Pal, ``Assertionbench: A benchmark to evaluate large-language models for assertion generation,'' \emph{arXiv preprint arXiv:2406.18627}, 2024.

\bibitem{10546729}
M.~Hassan, S.~Ahmadi-Pour, K.~Qayyum, C.~K. Jha, and R.~Drechsler, ``Llm-guided formal verification coupled with mutation testing,'' in \emph{2024 Design, Automation \& Test in Europe Conference \& Exhibition (DATE)}, 2024, pp. 1--2.

\end{thebibliography}

\end{document}